\documentclass[journal,final]{IEEEtran}
\usepackage[utf8]{inputenc}
\usepackage{amsmath,amssymb} \interdisplaylinepenalty=2500
\usepackage{graphicx} \graphicspath{{figs/}}
\usepackage{cite}
\usepackage{booktabs}

\newcommand{\SNR}{\textsf{SNR}}
\DeclareMathOperator{\trace}{\textsf{tr}}
\DeclareMathOperator{\rank}{\textsf{rank}}
\DeclareMathOperator{\diag}{\textrm{diag}}

\newcommand{\csum}{C_{\textrm{sum}}}
\newcommand{\csumhi}{C_{\textrm{sum}}^{\textrm{HI}}}
\newcommand{\rsum}{R_{\textrm{sum}}}

\newcommand{\rcomp}{R_{\textrm{comp}}}
\newcommand{\rif}{R_{\textrm{IF}}}

\newcommand{\rdif}{R_{\textrm{DIF}}}
\newcommand{\rdifhi}{R_{\textrm{DIF}}^{\textrm{HI}}}

\newcommand{\Dbar}{\bD_0}
\newcommand{\dbar}{\bar{d}}

\newcommand{\eff}{{\rm eff}}

\newtheorem{theorem}{Theorem}

\newtheorem{corollary}[theorem]{Corollary}

\def\qed{\endIEEEproof}

\newcommand{\mat}[1]{\begin{bmatrix} #1 \end{bmatrix}}

\newcommand{\ZZ}{\mathbb{Z}}

\newcommand{\RR}{\mathbb{R}}
\newcommand{\CC}{\mathbb{C}}

\renewcommand{\Re}{\mathsf{Re}}

\newcommand{\bzero}{\boldsymbol{0}}

\newcommand{\blambda}{\boldsymbol{\lambda}}

\newcommand{\calC}{\mathcal{C}}

\newcommand{\calN}{\mathcal{N}}

\newcommand{\calQ}{\mathcal{Q}}

\newcommand{\calV}{\mathcal{V}}
\newcommand{\calW}{\mathcal{W}}

\newcommand{\bA}{\mathbf{A}}

\newcommand{\bC}{\mathbf{C}}
\newcommand{\bD}{\mathbf{D}}

\newcommand{\bG}{\mathbf{G}}
\newcommand{\bH}{\mathbf{H}}
\newcommand{\bI}{\mathbf{I}}

\newcommand{\bM}{\mathbf{M}}

\newcommand{\bP}{\mathbf{P}}
\newcommand{\bQ}{\mathbf{Q}}

\newcommand{\bT}{\mathbf{T}}

\newcommand{\bW}{\mathbf{W}}
\newcommand{\bX}{\mathbf{X}}
\newcommand{\bY}{\mathbf{Y}}
\newcommand{\bZ}{\mathbf{Z}}

\newcommand{\ba}{\mathbf{a}}

\newcommand{\bc}{\mathbf{c}}
\newcommand{\bd}{\mathbf{d}}

\newcommand{\bh}{\mathbf{h}}

\newcommand{\bw}{\mathbf{w}}
\newcommand{\bx}{\mathbf{x}}
\newcommand{\by}{\mathbf{y}}
\newcommand{\bz}{\mathbf{z}}

\title{On Integer-Forcing Precoding for the Gaussian MIMO Broadcast Channel}
\author{Danilo~Silva,~\IEEEmembership{Member,~IEEE,}
		Gabriel~Pivaro, %~\IEEEmembership{Member,~IEEE,}
		Gustavo~Fraidenraich, %~\IEEEmembership{Member,~IEEE,}
		and Behnaam~Aazhang,~\IEEEmembership{Fellow,~IEEE}%
\thanks{Manuscript received November 23, 2016; revised February 22, 2017; accepted April 16, 2017. Date of publication MMMM DD, YYYY; date of current version MMMM DD, YYYY. The work of D.~Silva was supported in part by CNPq-Brazil under Grant 475482/2012-3. The work of G. F. Pivaro was supported in part by CAPES-Brazil under Grant BEX10714/14-6. The work of B. Aazhang is supported by NSF grants NeTS 1527811 and EARS 1547305. The associate editor coordinating the review of this paper and approving it for publication was N. Yang.}%
\thanks{D.~Silva is with the Department of Electrical and Electronic Engineering, Federal University of Santa Catarina, Florian\'{o}polis, SC 88040-900, Brazil (e-mail: danilo@eel.ufsc.br).}%
\thanks{G. Pivaro was with the Department of Communications, State University of Campinas, Campinas, SP 13083-852, Brazil. He is now with the Radiocommunications Reference Center, National Institute of Telecommunications, Santa Rita do Sapucai, MG 37540-000, Brazil (e-mail: gfp.1@hotmail.com).}%
\thanks{G. Fraidenraich is with the Department of Communications, State University of Campinas (Unicamp), Campinas, SP 13083-852, Brazil (e-mail: gf@decom.fee.unicamp.br).}
\thanks{B. Aazhang is with the Department of Electrical and Computer Engineering, Rice University, TX 77005, USA (e-mail: aaz@rice.edu).}%
\thanks{Color versions of one or more of the figures in this paper are available
online at http://ieeexplore.ieee.org.}%
\thanks{Digital Object Identifier XXXXX}%
\thanks{This work has been submitted to the IEEE for possible publication.  Copyright may be transferred without notice, after which this version may no longer be accessible.}%
}

\begin{document}

\maketitle

\begin{abstract}
Integer-forcing (IF) precoding, also known as downlink IF, is a promising new approach for communication over multiple-input multiple-output (MIMO) broadcast channels. Inspired by the integer-forcing linear receiver for multiple-access channels, it generalizes linear precoding by inducing an effective channel matrix that is approximately integer, rather than approximately identity.
Combined with lattice encoding and a pre-inversion of the channel matrix at the transmitter, the scheme has the potential to outperform any linear precoding scheme, despite enjoying similar low complexity.

In this paper, a specific IF precoding scheme, called \textit{diagonally-scaled exact IF} (DIF), is proposed and shown to achieve maximum spatial multiplexing gain. For the special case of two receivers, in the high SNR regime, an optimal choice of parameters is derived analytically, leading to an almost closed-form expression for the achievable sum rate. In particular, it is shown that the gap to the sum capacity is upper bounded by 0.27 bits for any channel realization. For general SNR, a regularized version of DIF (RDIF) is proposed. Numerical results for two receivers under Rayleigh fading show that RDIF can achieve performance superior to optimal linear precoding and very close to the sum capacity.

\end{abstract}


\section{Introduction}
\label{sec:introduction}

\IEEEPARstart{M}{ultiple-input}
multiple-output (MIMO) broadcast channels (BC) have received significant attention in recent years. One reason is that, provided that channel state information is available at the transmitter (CSIT), the sum rate of the system (also called throughput) scales linearly with the minimum of the number of transmit antennas and the aggregate number of receive antennas \cite{Tse.Viswanath}. Thus, given enough receivers, the throughput can be increased simply by increasing the number of transmit antennas, even if each receiver has only a single antenna \cite{Caire2003}.

While the capacity region of the Gaussian MIMO BC is well known, it has only been achieved so far using the so-called dirty-paper coding (DPC) technique \cite{Caire2003,Viswanath2003,Vishwanath.etal.2003.Duality-Achievable-Rates,Yu.Cioffi.2004.Sum-Capacity-Gaussian,Weingarten2006}.
However, DPC is widely regarded as being mostly of theoretical value, due to its significantly high implementation complexity \cite{Caire2003,Yoo.Goldsmith.2006.Optimality-Multiantenna-Broadcast}. As a consequence, suboptimal, lower-complexity schemes have been intensively investigated, with the goal of enabling a practical implementation that achieves good performance.

The simplest and yet very rich class of alternative methods is that of linear precoding (or beamforming), which consists of multiplying the (complex-valued) signals to be transmitted by a well-chosen matrix \cite{Yoo.Goldsmith.2006.Optimality-Multiantenna-Broadcast,Bjornson2014}, aiming at producing parallel channels to each receiver, while balancing residual interference and noise. Popular methods in this class include zero-forcing (ZF) precoding and regularized zero-forcing (RZF) precoding.

While these methods are simple and achieve the same multiplexing gain as DPC, they suffer a significant penalty when the aggregate number of receive antennas is equal or slightly less than the number of transmit antennas, even for asymptotically high SNR \cite{Lee2007}. However, even \textit{optimal} linear precoding \cite{Bjornson2014} (which is currently infeasible to implement in practice) falls short of achieving the DPC performance.

\subsection{Integer-Forcing Precoding}

Recently, an integer-forcing (IF) approach to the problem has been proposed that appears to provide promising gains without a significant increase in complexity \cite{Hong2012,Hong2013,He2014,He.etal.2014.Uplink-downlink-Duality-Integer-forcing:Iterative-Optimization}. The approach is inspired by the compute-and-forward (CF) framework for the multiple-access channel (MAC) \cite{Nazer2011} and can be understood as the dual of the integer-forcing \textit{equalization} approach for the MIMO MAC \cite{Zhan2014}. Fundamentally, it is a nonlinear technique that nevertheless enjoys many of the properties of traditional linear schemes.

From a high level, the IF approach to precoding can be described as a concatenation of three layers. In the inner layer (closest to the channel), the transmitter applies linear beamforming so that the precoded channel matrix becomes approximately an integer matrix $\bA$; this is referred to as precoding at the signal level, or \textit{signal precoding}. In the outer layer, the transmitter precodes the original messages into auxiliary messages, pre-inverting the integer channel matrix $\bA$ that will be seen by the receivers. This is referred to precoding at the message level, or \textit{message precoding}, and is entirely performed using finite-field operations. In the middle layer, the transmitter channel encodes the auxiliary messages using nested lattice codes, forming the codewords to which signal precoding is applied. Each receiver then attempts to decode an integer linear combination of these codewords, where the coefficients of the linear combination correspond to the rows of the integer matrix $\bA$. Due to the integer channel pre-inversion, each correctly decoded message is exactly the original message intended to that receiver. Moreover, assuming that lattice encoding and decoding can be done efficiently, the overall complexity is only slightly higher than that of classical linear precoding, essentially due to the message precoding stage.

\subsection{Related Work}

The original publication \cite{Zhan2014} and most of the subsequent work on IF refers to IF \textit{equalization} for a multi-antenna receiver. More precisely, it is a receiver architecture since it involves three layers: linear equalization, channel decoding and integer channel inversion (which also can be understood as message-level equalization). Essentially, the equalization task is broken in two parts, shifting the channel decoding task to the middle. The name ``integer forcing'' refers to the goal of the equalizer, which is to produce not necessarily \textit{zero} interference, as in zero-forcing equalization, but \textit{integer}-scaled interference, which can then be easily unmixed after noise has been removed. The fundamental ingredient for this architecture, which enables the decoder to work as expected even under integer-scaled interference, is the use of lattice codes with lattice decoding. Since these are known to be capacity-achieving for the Gaussian channel, the rates achieved are never inferior (and are usually much higher) than those achieved by conventional equalization.

While the IF receiver architecture was originally conceived for a multiuser uplink scenario or a single-user MIMO system with independent-stream (V-BLAST) encoding, it can of course also be applied in single-user MIMO with joint processing at the transmitter, which can only improve its performance. For instance, one interesting modification is the introduction of linear precoding on the independently-encoded streams. This approach is known as \textit{precoded integer-forcing} (equalization) and has been optimized to approach the MIMO capacity \cite{Ordentlich.Erez.2015.Precoded-Integer-Forcing-Universally} or to achieve full diversity \cite{Sakzad.Viterbo.2015.Full-Diversity-Unitary}.

In contrast, IF precoding is a transmission architecture for a multi-antenna transmitter that may be applied either to a downlink multiuser system---in which case it may also be referred to as \textit{downlink} IF---or to a single-user MIMO system with separate processing at each receive antenna. Thus, as the uplink and downlink IF problems are fundamentally different, techniques originally developed for the former are not necessarily applicable to the latter, although under certain assumptions a duality relation can be established between the two \cite{He2014}. To the best of our knowledge, downlink IF has only been previously considered in \cite{Hong2012,Hong2013,He2014,He.etal.2014.Uplink-downlink-Duality-Integer-forcing:Iterative-Optimization}.

Originally, a downlink IF scheme without beamforming was proposed in \cite{Hong2012,Hong2013}, where it was named reverse compute-and-forward (RCF). A more general version allowing beamforming was also proposed in \cite{Hong2012}; however, a specific choice of the beamforming matrix was used, which enforced the precoded channel matrix to be exactly integer. The most general form of downlink IF, allowing arbitrary linear beamforming, as well as potentially different shaping lattices for each user,
was later proposed in \cite{He2014} in the context of an uplink-downlink duality result for IF. A disadvantage of such a high degree of generality is that the problem of finding an optimal choice of parameters---which include not only the beamforming and integer matrices, but also a power vector specifying the second moment of each shaping lattice---becomes very involved. An iterative optimization algorithm, based on the uplink-downlink duality, is proposed in \cite{He.etal.2014.Uplink-downlink-Duality-Integer-forcing:Iterative-Optimization}. However, this algorithm is not guaranteed to find an optimal solution and requires a relatively computationally expensive lattice basis reduction step at each iteration; moreover, as an arbitrary power vector may result from the optimization, the algorithm may not be applicable to schemes that use a single shaping lattice for all users.

It is worth mentioning that IF techniques are closely related to lattice reduction strategies \cite{Yao.Wornell.2002.Lattice-reduction-aided-Detectors-MIMO,Fischer.etal.2016.Factorization-Approaches-Lattice-Reduction-Aided}. It is shown in \cite{Zhan2014} that uplink IF with uncoded transmission, i.e., using a one-dimensional lattice, is essentially equivalent to lattice-reduction-aided reception, which focuses on symbol-by-symbol detection. In contrast, IF with a sufficiently long code can guarantee reliable communication for any (fixed) channel realization, therefore providing an achievable rate. Exactly the same observations can be made for downlink IF and its relation to lattice-reduction-aided precoding/preequalization \cite{Windpassinger.etal.2004.Lattice-reduction-aided-Broadcast-Precoding}, which is further explored in \cite{Stern.Fischer.2016.Advanced-Factorization-Strategies}.

\subsection{Outline of This Work}

In this work, we restrict our attention to the special case of downlink IF with a single shaping lattice for all users and focus on the optimization of the signal precoding and integer matrices.

We propose a specific structure for the precoding matrix called \textit{diagonally-scaled exact integer-forcing} (DIF). This scheme is a generalization of the exact IF scheme in [12] by allowing arbitrary scaling by a diagonal matrix, akin to the power scaling in conventional beamforming. As we shall see, such a \textit{moderate} degree of generalization is key to obtaining a scheme that is flexible yet amenable to optimization.

We analyze the performance of DIF for high SNR and show that it achieves maximum spatial multiplexing gain. For the specific case of two receivers and high SNR, we derive the optimal solution for both the precoding matrix and the integer matrix. In a surprising contrast to IF for the MAC case---which typically requires solving a lattice basis reduction problem in order to find an optimal integer matrix---we show that both matrices can be found analytically and therefore very efficiently. This result in turn provides an expression for the achievable sum rate of DIF in almost closed form. In particular, we show that the gap between the sum rate achievable by DIF and that of DPC is upper bounded by 0.27 bits for any channel matrix. For general SNR, we propose a regularized version of DIF (RDIF) and show numerically that the method achieves a sum-rate performance very close to DPC.

In order to keep the paper self-contained, we opted to provide a complete description of a generic downlink IF scheme, comprising lattice construction, encoding and decoding, along with a simple proof that each receiver can indeed correctly compute its desired message from its corresponding observation. While this full picture can be understood by combining the results of \cite{He2014} and \cite{Hong2013}, as suggested in \cite{He.etal.2014.Uplink-downlink-Duality-Integer-forcing:Iterative-Optimization}, we believe that a concise description under a single notation may be more accessible for a reader unfamiliar with these previous works.

The remainder of this paper is organized as follows. In Section~\ref{sec:prelim}, the system model is described and the necessary background on downlink IF is reviewed. The main optimization problem is stated in Section~\ref{sec:problem-statement}, along with connections to related problems considered in previous works. In Section~\ref{sec:exact-if}, we describe our proposed scheme and analyze its performance in the high SNR regime. In Section~\ref{sec:two-user}, we focus on the two receiver case and derive the optimal selection of the precoding and integer matrices in high SNR regime. Moreover, we show in this section that the gap for the sum capacity in this scenario is bounded. Section~\ref{sec:General SNR} extends the scheme for the general SNR case. Finally, in Section~\ref{sec:Simulation Results}, we present numerical results on the average sum rate and the average gap to capacity of the proposed schemes under Rayleigh fading.

\section{Preliminaries}
\label{sec:prelim}

\subsection{Notation}

For any $x > 0$, let $\log_2^+(x)=\max\{0,\log_2(x)\}$. 
Unless specified otherwise, vectors are defined as row vectors.
For any vector~$\ba$, $\|\ba\|$ denotes its Euclidean norm. For any matrix~$\bA$, $\bA^{H}$ denotes its conjugate transpose and $\trace(\bA)$ denotes its trace. The identify matrix is denoted by $\mathbf{I}$. 

Let $\ZZ[j] = \ZZ + j\ZZ$ denote the ring of Gaussian integers. For any prime $p \in \ZZ$, let $\ZZ_p$ denote the ring of integers modulo~$p$ and let $\ZZ_p[j] = \ZZ_p + j\ZZ_p \cong \ZZ[j]/p\ZZ[j]$ denote the ring of Gaussian integers modulo~$p$. For $x \in \ZZ[j]$, we use $x \bmod p$ to denote the modulo-$p$ reduction of both the real and imaginary parts of $x$, i.e., $x \bmod p \in \{a+jb: a,b \in \{0,1,\ldots,p-1\}\}$. 

Whenever scalar functions are applied to vectors and matrices, it should be understood that the functions are applied element-wise.

\subsection{Lattices}

We review a few basic concepts on lattices, which can be found in more detail in, e.g., \cite{Zamir.2014.Lattice-Coding-Signals}.

A lattice $\Lambda \in \RR^n$ is a discrete $\ZZ$-submodule of $\RR^n$, i.e., it is 
closed to integer linear combinations \cite{Zamir.2014.Lattice-Coding-Signals}. More generally, a $\ZZ[j]$-lattice $\Lambda \in \CC^n$ is a discrete $\ZZ[j]$-submodule of $\CC^n$, i.e., it is closed to Gaussian integer linear combinations. 

The nearest neighbor quantizer $\calQ_\Lambda: \CC^n \to \Lambda$ is defined by $\bx \mapsto \arg\min_{\blambda \in \Lambda}\, \|\bx - \blambda\|$, with ties broken in a systematic manner. The Voronoi region of $\Lambda$ is defined as $\calV_\Lambda \triangleq \{\bx~\in~\CC^n: \calQ_\Lambda(\bx) = \bzero\}$ and the modulo-$\Lambda$ operation is defined as $\bx \bmod \Lambda \triangleq \bx - \calQ_\Lambda(\bx) \in \calV_\Lambda$. Note that $\bx \bmod \Lambda = \by$ implies that $\bx - \by \in \Lambda$. The second moment (per dimension) of $\Lambda$ is defined as $P_{\Lambda} = \frac{1}{n}\mathbb{E}[\|\bx\|^2]$, where $\bx$ is a random vector uniformly distributed over $\calV_{\Lambda}$.

If $\Lambda$ and $\Lambda_s \subseteq \Lambda$ are lattices, then $\calC = \Lambda \cap \calV_{\Lambda_s}$ is said to be a \textit{nested lattice code(book)}. Note that $\calC = \Lambda \bmod \Lambda_s$.

\subsection{System Model}
\label{ssec:system-model}

Consider the discrete-time complex baseband model of a Gaussian MIMO BC with one transmitter and $K$ receivers, where the transmitter has $M \geq K$ antennas and each receiver has a single antenna. 

Let $\bx_j' \in \CC^n$ be the vector sent by the transmitter on its $j$th antenna, $j=1,\ldots,M$, over $n$ channel uses. For $i=1,\ldots,K$, the vector received by the $i$th receiver is given by
\begin{equation}\label{eq:rx:sig}
 \by_i = \sum_{j=1}^M h_{ij} \bx_j' + \bz_i
\end{equation}
where $\bh_i = \mat{h_{i1} & \cdots & h_{iM}} \in \CC^{1 \times M}$ is the vector of channel coefficients and $\bz_{i}$ is a circularly symmetric jointly-Gaussian complex random vector with i.i.d. components $\sim \calC\calN(0,1)$.
Equivalently, we can write
\begin{equation}
 \bY = \bH \bX' + \bZ
\label{eq:channel:equation}
\end{equation}
where $\bY \in \CC^{K \times n}$, $\bH \in \CC^{K \times M}$, $\bX' \in \CC^{M \times n}$, and $\bZ \in \CC^{K \times n}$ are matrices having the vectors $\by_i$, $\bh_i$, $\bx_j'$ and $\bz_i$, respectively, as rows.

The transmit signals must satisfy an average total power constraint
\begin{equation}
\label{eq:power-constraint-total}
\frac{1}{n}\mathbb{E}[\trace(\bX'\bX'^H)] = \frac{1}{n}\sum_{j=1}^M \mathbb{E}\left[\|\bx'_{j}\|^2\right] \leq \SNR
\end{equation}
which is denoted by SNR since the noise is assumed to have unit variance.
We assume that the transmitter and each receiver have perfect knowledge of their respective channel coefficients.
We also assume that $\bH$ is full-rank.

We consider the problem where, for $i=1,\ldots,K$, an independent message $\bw_i \in \calW_i$, of rate $R_i = \frac{1}{n} \log_2 |\calW_i|$, taken from a message space $\calW_i$, is to be transmitted to the $i$th receiver. The sum rate of the scheme is given by $\rsum = R_1 + \cdots + R_K$.

A sum rate~$R$ is said to be achievable if, for any $\epsilon > 0$ and sufficiently large~$n$, there exists a coding scheme with sum rate~$R$ that allows each receiver to recover its intended message with probability of error smaller than $\epsilon$. The \emph{sum capacity} is the supremum of all achievable sum rates. It is well-known that the sum capacity of the MIMO BC is given by \cite{Viswanath2003,Vishwanath.etal.2003.Duality-Achievable-Rates,Yu.Cioffi.2004.Sum-Capacity-Gaussian,Weingarten2006}
\begin{equation}
\csum = \sup_{\bQ: \trace(\bQ) \leq 1} \log_2 \det\left(\bI + \SNR\ \bH^H \bQ \bH \right)
 \label{eq:sum:cap}
\end{equation}
where $\bQ$ is a $K \times K$ diagonal matrix.

It is also useful to define the spatial multiplexing gain of the scheme as
\begin{equation}
\lim_{\SNR\to \infty}\frac{\rsum}{\log_2(\SNR)}.
\end{equation}
It can be shown that the maximum spatial multiplexing gain achievable is equal to~$K$.

\subsection{Integer-Forcing Precoding}
\label{sec:integer-forcing}

We now review the IF precoding approach to the MIMO~BC, originally proposed in \cite{Hong2012} and extended in \cite{He2014}. The scheme consists of the steps of message precoding, nested lattice encoding, and linear beamforming, performed at the transmitter, together with compute-and-forward (lattice) decoding at each receiver.

We restrict our attention to the case of IF with symmetric coding power allocation, where a single shaping lattice is used for all receivers, which is enough for the purposes of this paper. This is a special case of the scheme in \cite{He2014}, which allows multiple shaping lattices with unequal second moments. Such a special case, while potentially inferior to \cite{He2014} in terms of achievable rates, has the advantage of being easier to optimize and requiring lower implementation complexity, which is relevant if IF is to be competitive against conventional low-complexity beamforming schemes.

\subsubsection{Construction}

Let $p \in \ZZ$ be a prime satisfying $p - 3 \in 4\ZZ$, so that $\ZZ_p[j]$ becomes a finite field \cite{Dummit.Foote}.

For $i=1,\ldots,K$, let the message space for the $i$th user be given by $\calW_i = \ZZ_p[j]^{k_i} \times \{0\}^{n-k_i} \subseteq \calW$, where $\calW = \ZZ_p[j]^n$ is the \textit{ambient space}. In other words, the elements of~$\calW_i$ are length-$n$ (row) vectors over $\ZZ_p[j]$ whose last $n-k_i$ entries are zeros.

Let $\Lambda \subseteq \CC^n$ be a $\ZZ[j]$-lattice, referred to as the \textit{fine} or \textit{coding} lattice, and let $\Lambda_s = p\Lambda$, referred to as the \textit{coarse} or \textit{shaping} lattice. The scaling for $\Lambda$ is chosen so that $P_{\Lambda_s} = \SNR$. Let $\calC = \Lambda \cap \calV_{\Lambda_s}$ be a nested lattice code, which is to be used as a ``mother'' code for the transmitter. 
Let $\varphi: \Lambda \to \calW$ be a surjective $\ZZ[j]$-linear map\footnote{More precisely, a $\ZZ[j]$-module homomorphism.} with kernel $\Lambda_s$ (referred to as a linear labeling in \cite{Feng2013}) and let $\tilde{\varphi}: \calW \to \calC$ be a bijective encoding function such that $\varphi(\tilde{\varphi}(\bw)) = \bw$, for all $\bw \in \calW$. For all $i$, define the lattice code $\calC_i = \tilde{\varphi}(\calW_i) \subseteq \calC$ as the image of the encoder restricted to $\calW_i$, which naturally results in a subcode of $\calC$. 

Each $i$th receiver is assumed to have a lattice decoder for the corresponding $\calC_i$, while the transmitter is assumed to have an encoder for $\calC$.

\subsubsection{Encoding}

Let $\bW \in \calW^{K} = \ZZ_p[j]^{K \times n}$ be a matrix whose rows are the messages $\bw_1,\ldots,\bw_K$. Let $\bA \in \ZZ[j]^{K \times K}$ be an integer matrix that is invertible modulo $p$, i.e., for which
\begin{equation}\label{eq:determinant-modulo-constraint}
 \det(\bA) \bmod p \neq 0
\end{equation}
and let $\tilde{\bA} \in \ZZ[j]^{K \times K}$ be any matrix satisfying
\begin{equation}
\bA\tilde{\bA} \bmod p = \bI.
\end{equation}

First, the original messages are linearly precoded with $\tilde{\bA}$, resulting in the precoded messages $\bw_1',\ldots,\bw_K' \in \calW$ given as the rows of the matrix
\begin{equation}
\bW' = \tilde{\bA} \bW.
\end{equation}

Then, each precoded message $\bw_i'$ is encoded with the lattice code $\calC$, yielding a codeword $\bc_i' = \tilde{\varphi}(\bw_i') \in \calC$. Next, for each~$i$, a vector
\begin{equation}
\bx_i = \bc_i' + \bd_i \bmod \Lambda_s
\end{equation}
is computed, where $\bd_i \in \CC^n$ is a dither vector, selected\footnote{For instance, $\bd_i$ may be chosen to be uniformly distributed over $\calV_{\Lambda_s}$, but fixed dithers are also possible \cite{Nazer.etal.2016.Expanding-Compute-and-Forward-Framework:}.} so as to ensure that 
$\frac{1}{n}\mathbb{E}\left[\|\bx_{i}\|^2\right] \leq P_{\Lambda_s} = \SNR$.

Finally, linear beamforming---also referred to here as \textit{signal} precoding\footnote{The expressions \textit{signal precoding} and \textit{beamforming} are used interchangeably in this paper.}---with a matrix $\bT \in \CC^{M \times K}$ is performed, producing
the transmission matrix 
\begin{equation}
 \bX' = \bT \bX
\end{equation}
where $\bX \in \CC^{K \times n}$ is the matrix whose rows are $\bx_1,\ldots,\bx_K$.
Note that the beamforming matrix must satisfy the constraint
\begin{equation}
\label{eq:precoding-matrix-constraint}
\trace(\bT^H\bT) \leq 1
\end{equation}
in order to guarantee that the power constraint \eqref{eq:power-constraint-total} is respected.

\subsubsection{Decoding}

Let $\bC', \bD \in \CC^{K \times n}$ be matrices whose rows are, respectively, the codewords $\bc_1',\ldots,\bc_K'$ and the dither vectors $\bd_1,\ldots,\bd_K$, and let $\ba_1,\ldots,\ba_K \in \ZZ[j]^K$  denote the rows of~$\bA$.\footnote{In practice, the matrix $\bA$ chosen by the transmitter would have to be communicated to the receivers prior to the main transmission, possibly in a preamble section of the transmitted frame.}

Recall that the observation of the $i$th receiver is given by
\begin{equation}
\by_i = \bh_i \bX' + \bz_i = \bh_i' \bX + \bz_i
\end{equation}
where $\bh_i' = \bh_i \bT$. The receiver selects a scalar $\alpha_i \in \CC$ and computes
\begin{align}
\by_{\eff,i} 
&= \alpha_i \by_i - \ba_i \bD \bmod \Lambda_s \\
&= \alpha_i (\bh_i' \bX + \bz_i) - \ba_i \bD \bmod \Lambda_s \\
&= \ba_i (\bX - \bD)  + \underbrace{(\alpha_i \bh_i' - \ba_i) \bX + \alpha_i \bz_i}_{\bz_{\eff,i}} \bmod \Lambda_s \\
&= \ba_i \bC'  + \bz_{\eff,i} \bmod \Lambda_s \\
&= \bc_i + \bz_{\eff,i} \bmod \Lambda_s \label{eq:effective-channel}
\end{align}
where $\bc_i = \ba_i \bC' \bmod \Lambda_s$ and
\begin{equation}
\bz_{\eff,i} = (\alpha_i \bh_i' - \ba_i) \bX + \alpha_i \bz_i
\end{equation}
is the effective noise.

Since $\varphi$ is $\ZZ[j]$-linear with kernel $\Lambda_s$, we have that
\begin{equation*}
\varphi(\ba_i \bC' \bmod \Lambda_s) = \ba_i\varphi(\bC') = \ba_i\bW' = \ba_i \tilde{\bA} \bW = \bw_i
\end{equation*}
which implies that $\bc_i = \tilde{\varphi}(\bw_i) \in \calC_i$, i.e., $\bc_i$ corresponds exactly to the encoding of the original message $\bw_i$, even though it is never explicitly computed at the transmitter.

It follows that, if the receiver can correctly decode $\bc_i$ from $\by_{\eff,i}$, rejecting the effective noise $\bz_{\eff,i}$, then the message $\bw_i$ can be easily recovered by inverting $\tilde{\varphi}$.

\subsubsection{Achievable rates}

According to \eqref{eq:effective-channel}, the $i$th receiver effectively sees an independent modulo-$\Lambda_s$ channel \cite{Erez.Zamir.2004.Achieving-1/2-Log}, free of interference from other messages (except as already incorporated in $\bz_{\eff,i}$).
With an appropriate lattice construction, it can be shown \cite{Nazer2011} that, provided that $k_1,\ldots,k_K$ and $p$ are allowed to grow with $n$, the following rate tuple is achievable:
\begin{equation}
R_i = \log^+\left(\frac{\SNR}{\sigma_{\eff,i}^2}\right), \quad i=1,\ldots,K
\end{equation}
where
\begin{equation}
\sigma_{\eff,i}^2 = \frac{1}{n} \mathbb{E}\left[\| \bz_{\eff,i} \|^2\right]
\end{equation}
is the per-component variance of the effective noise.

After optimizing each scalar $\alpha_i$, this achievable rate tuple can be expressed more simply as
\begin{equation}
R_i = \rcomp(\bh_i',\ba_i), \quad i=1,\ldots,K
\end{equation}
where
\begin{equation}\label{eq:comprate-original}
\rcomp(\bh_i',\ba_i) \triangleq \log_2^+\left(\frac{1}{
\ba_i \left(\bI - \frac{\SNR}{\SNR \|\bh_i'\|^2 + 1}\bh_i'^H \bh_i'\right) \ba_i^H}\right)
\end{equation}
is known as a \textit{computation rate} in the compute-and-forward framework \cite{Nazer2011}.

It follows that the sum rate achievable by the IF scheme is given by
\begin{equation}
 \rif(\bA,\bT) \triangleq \sum_{i=1}^K \rcomp(\bh_i \bT,\ba_i) = \sum_{i=1}^K \rcomp(\bh_i',\ba_i)
\end{equation}
where, for conciseness, we omit the dependence on $\bH$ and $\SNR$.

\section{Problem Statement}
\label{sec:problem-statement}

In this paper, we focus on the problem of choosing $\bA$ and $\bT$, based on $\bH$ and $\SNR$, in order to maximize the achievable sum rate $\rif(\bA,\bT)$. More precisely, the maximum achievable sum rate for the scheme is defined as
\begin{equation}\label{eq:sum-rate-optimized-full}
\rif \triangleq \max_{\substack{\bA \in \ZZ[j]^{K \times K}, \bT \in \CC^{M \times K}:\\ \rank \bA = K\\ \trace(\bT^H \bT) \leq 1}}\; \rif(\bA,\bT).
\end{equation}

Note that, in the above expression, $\bA$ is only required to be invertible over $\CC$, rather than modulo $p$. This is due to the fact that, without loss of optimality, we may safely restrict the search to matrices $\bA$ of bounded determinant,%
\footnote{This follows provided that $\|\bh_i\|^2$ is bounded, which implies that both $\|\bh_i'\|^2$ and $\|\ba_i\|^2$ are bounded as well. See \cite[Remark 10]{Nazer2011} for details.
%Proof that $\|\bh_i \bT\|^2 \leq \|\bh_i\|^2$:
%\begin{align}
%\|\bh_i \bT\|^2 
%&\leq \|\bh_i\|^2 \lambda_{\max}(\bT\bT^H) \nonumber \\
%&\leq \|\bh_i\|^2 \trace(\bT\bT^H) \nonumber \\
%&\leq \|\bh_i\|^2 \nonumber
%\end{align}
%where $\lambda_{\max}(\bT\bT^H)$ denotes the maximum eigenvalue of $\bT\bT^H$.
}
while, as far as we are concerned with achievable rates, the modulus $p$ has to grow without bound, implying that any optimal matrix for (\ref{eq:sum-rate-optimized-over-coefficient-matrix}) will also satisfy (\ref{eq:determinant-modulo-constraint}). In other words, the modulo $p$ issue is irrelevant for achievable rates, although it may be important in a practical implementation.

Without any loss of optimality, we can always decompose \eqref{eq:sum-rate-optimized-full} as
\begin{equation}\label{eq:sum-rate-optimized-over-coefficient-matrix}
 \rif = \max_{\bA: \rank \bA = K}\;  \rif(\bA)
\end{equation}
where
\begin{equation}\label{eq:sum-rate-optimized-precoding}
 \rif(\bA) \triangleq \max_{\bT: \trace(\bT^H \bT) \leq 1}\; \rif(\bA,\bT).
\end{equation}
Thus, we can focus on solving the continuous problem \eqref{eq:sum-rate-optimized-precoding} for all $\bA$ before attempting the integer problem \eqref{eq:sum-rate-optimized-over-coefficient-matrix}. 

In particular, if we take $\bA = \bI$, then \eqref{eq:sum-rate-optimized-precoding} becomes equivalent to optimal linear precoding \cite{Bjornson2014} with the sum rate as objective. This is due to the fact that, in this case, the computation rate $\rcomp(\bh_i \bT,\ba_i)$ becomes equal to the capacity of the channel to the $i$th receiver when treating interference from all other users as noise.

Unfortunately, optimal linear precoding is known to be NP-hard in general (see \cite{Bjornson2014} and references therein), so \eqref{eq:sum-rate-optimized-full} is likely to be also NP-hard, as it certainly is at least in all channel realizations where the choice $\bA= \bI$ is optimal. Thus, except for the special case of high SNR, we will not attempt at an optimal solution but rather at a low-complexity approach that yields good performance. Note that, in principle, problem \eqref{eq:sum-rate-optimized-full} has to be solved at the transmitter for every channel realization, so an efficient implementation is essential.

\subsection{Related Schemes}

When $M=K$ and $\bT = \bI$, the scheme reduces to the original RCF \cite{Hong2013}. In this case, a natural strategy for optimizing $\bA$ is to choose each coefficient vector $\ba_i$  independently for each receiver as the one which maximizes (\ref{eq:comprate-original}) for its corresponding channel gain vector~$\bh_i$. This approach is optimal (for this specific $\bT$) if the resulting matrix $\bA$ turns out to be full rank, but this condition is not guaranteed to occur. This problem is avoided in \cite{Hong2013} by restricting the transmission to a subset of receivers (and antennas) for which the full rank condition is obtained. However, this solution is not applicable when the number of receivers is fixed, which is the scenario considered here.
 
When $M=K$ and $\bT = c \bH^{-1} \bA$, where $c > 0$ is chosen to satisfy a per-antenna power-constraint, then the scheme reduces to RCF with ``integer-forcing beamforming'' proposed in \cite{Hong2012}, which produces an exactly integer effective channel matrix. In this case it is no longer possible to optimize each $\ba_i$ individually, since $\bT$ (and thus $\bh_i' = \bh_i \bT$) depends on $\bA$. Therefore, the optimization of $\bA$ must be based on the complete sum rate $\rif(\bA,\bT)$. This is a hard problem in general, suggesting the use of suboptimal methods, such as the one proposed in \cite{Hong2012}.

However, as we shall see, having more flexibility on the choice of $\bT$ actually makes the problem more structured and potentially easier to solve.

It is also worth mentioning that if we take $\bA = \bI$ and $\bT = \bH^H {(\bH\bH^H)}^{-1}\bD$, where $\bD$ is a diagonal matrix, we recover ZF precoding \cite{Bjornson2014}, while if we take $\bT = \bH^H \left(\frac{K}{\SNR}\bI + \bH\bH^H\right)^{-1} \bD$, we recover RZF precoding \cite{Bjornson2014}.

\section{Proposed Scheme}
\label{sec:exact-if}

\subsection{Optimal Precoding Structure for High SNR}

In this section we propose and analyze a signal precoding structure that is the main focus of this paper. Our definition is motivated by the following result.

\medskip
\begin{theorem}\label{thm:spatial-multiplexing}
For any full-rank $\bA$, integer-forcing precoding with a fixed $\bT$ achieves maximum spatial multiplexing gain if and only if
\begin{equation}\label{eq:eif-condition}
\bH \bT = \bD \bA
\end{equation}
for some diagonal matrix 
$\bD = \diag(d_1,\ldots,d_K)$ with 
nonzero diagonal entries. In this case,
\begin{equation}\label{eq:exact-sum-rate-high-snr}
\rif(\bA,\bT) =
\sum_{i=1}^K \log_2^+\left(\frac{1}{\|\ba_i\|^2}  + |d_i|^2 \SNR \right)
\end{equation}
and
\begin{equation}
\lim_{\SNR\to \infty}\frac{\rif(\bA,\bT)}{\log_2(\SNR)} = K.
\end{equation}

\end{theorem}
\begin{IEEEproof}
The computation rate (\ref{eq:comprate-original}) for the $i$th receiver can be rewritten as
\begin{align}
&\rcomp(\bh_i',\ba_i) \nonumber \\
&= \log_2^+\left(\frac{1}{
\ba_i \left(\bI - \frac{\SNR}{\SNR \|\bh_i'\|^2 + 1}\bh_i'^H \bh_i'\right) \ba_i^H}\right) \nonumber \\
&= \log_2^+\left(\frac{1 + \|\bh_i'\|^2 \SNR}{
\ba_i \left[(1 + \|\bh_i'\|^2 \SNR) \bI - \SNR \bh_i'^H \bh_i'\right] \ba_i^H}\right) \nonumber \\
&= \log_2^+\left(\frac{1 + \|\bh_i'\|^2\SNR}{\|\ba_i\|^2 + (\|\ba_i\|^2 \|\bh_i'\|^2 - |\bh_i' \ba_i^H |^2)\SNR}\right). \label{Rcomp:eq}
\end{align}

From the Cauchy-Schwarz inequality, we know that the term
\begin{equation}
 \|\ba_i\|^2 \|\bh_i'\|^2 - |\bh_i'\ba_i^H|^2
\end{equation}
is always non-negative, and it is equal to zero if and only if $\bh_i' = d_i \ba_i$, for some $d_i \in \CC$. Thus, for fixed $\bh_i',\ba_i$, the computation rate grows with SNR if and only if $\bh_i' \neq 0$ and $\bh_i' = d_i \ba_i$. In this case, we obtain
\begin{equation}
\rcomp(\bh_i',\ba_i) = \log_2^+\left(\frac{1}{\|\ba_i\|^2}  + |d_i|^2 \SNR \right).
\end{equation}
The remaining statements follow directly.
\end{IEEEproof}
\medskip

Theorem~\ref{thm:spatial-multiplexing} shows that the optimal IF precoder structure for high SNR is such that the precoded channel matrix becomes exactly an integer matrix, up to scaling for each user. We call any such scheme \textit{diagonally-scaled exact integer-forcing} (DIF) precoding. Clearly, DIF generalizes ZF, which corresponds to the special case $\bA = \bI$. It also generalizes the ``exact IF'' scheme in \cite{Hong2012}, which corresponds to the special case $\bD = c\bI$.

Note that the DIF precoding structure is optimal for high SNR in the sense that it contains all of the optimal solutions to \eqref{eq:sum-rate-optimized-full} when $\SNR \to \infty$, i.e., any choice of $(\bA,\bT)$ that violates this structure cannot be optimal in this regime. In particular, achieving maximum spatial multiplexing gain is a necessary condition for a bounded gap to capacity. Of course, an optimal precoding structure for finite SNR must depend on the SNR (so $\bT$ is not fixed), in such a way that it converges to DIF as SNR grows.

For any $\bA$ and $\bD = \diag(d_1,\ldots,d_K)$, let
\begin{equation}
\rdif(\bA,\bD) \triangleq \sum_{i=1}^K \log_2^+\left(\frac{1}{\|\ba_i\|^2}  + |d_i|^2 \SNR \right)
\end{equation}
\begin{equation}
\begin{array}{rll}
\rdif(\bA) \triangleq & \max\limits	_{\bD,\bT} & \rdif(\bA,\bD) \\
    & \text{s.t.} & \bH \bT = \bD \bA \\
    &             & \trace(\bT \bT^H) \leq 1
\end{array}
\end{equation}
and
\begin{equation}
\rdif \triangleq \max_{\bA: \rank \bA = K}\; \rdif(\bA).
\end{equation}

The question of optimally choosing $\bT$ given $\bA$ and $\bD$ under a power constraint is addressed in the following theorem.

\medskip
\begin{theorem}\label{thm:precoder-structure-optimal}
For any full-rank $\bA$,
\begin{equation}\label{eq:precoder-optimization-problem}
\begin{array}{rll}
\rdif(\bA) = & \max\limits	_{\bD} & \rdif(\bA,\bD) \\
    & \text{s.t.} & \trace(\bT \bT^H) \leq 1
\end{array}
\end{equation}
with $\bT$ given by
\begin{equation}\label{T:eq}
\bT = \bH^H (\bH\bH^H)^{-1} \bD \bA.
\end{equation}
\end{theorem}
\begin{IEEEproof}
The proof is a direct generalization of that of \cite[Theorem 1]{Wiesel2008} and is therefore omitted.
\end{IEEEproof}
\medskip

Note that, for $\bA = \bI$, problem \eqref{eq:precoder-optimization-problem} is equivalent to the conventional power allocation problem for ZF precoding, which can be expressed as a simple concave maximization with one linear constraint \cite{Wiesel2008}. However, for general $\bA$, problem \eqref{eq:precoder-optimization-problem} does not appear to be concave, due to both the objective function and the constraint.

\subsection{Equivalent Objective Function for High SNR}

We are particularly interested in analyzing the DIF scheme in the high SNR regime.
For any $\bA$ and diagonal $\bD$, let
\begin{equation}
\rdifhi(\bA,\bD) \triangleq \log_2\left(|\det\bD|^2 \SNR^K \right)
\end{equation}
and
\begin{equation*}
\rdifhi(\bA) \triangleq \max_{\bD}\, \rdifhi(\bA,\bD)
\end{equation*}
subject to the same constraint as in Theorem~\ref{thm:precoder-structure-optimal}, and let
\begin{equation*}
\rdifhi \triangleq \max_{\bA}\, \rdifhi(\bA).
\end{equation*}

It is easy to see that $\rdif \geq \rdifhi$ and
\begin{equation*}
\lim_{\SNR \to \infty} \rdif - \rdifhi = 0.
\end{equation*}
Thus, for high SNR, $\rdifhi(\bA,\bD)$ can serve as an equivalent objective function that is easier to analyze than $\rdif(\bA,\bD)$. We will focus on this function for the remainder of the paper.

\medskip
\begin{theorem}\label{thm:rate-hi-snr-objfun}
For any full-rank $\bA$,
\begin{equation}\label{eq:rate-hi-snr-objfun}
\rdifhi(\bA) = \max\limits_{\Dbar: |\det\Dbar|=1}\;  K \log_2\left(\frac{\SNR}{\trace(\bT_0 \bT_0^H)} \right) \\
\end{equation}
where $\Dbar$ is a diagonal matrix and
\begin{equation}\label{eq:aux-precoding-matrix}
\bT_0 = \bH^H (\bH\bH^H)^{-1} \Dbar \bA.
\end{equation}
This value is achievable by choosing $\bD = c \Dbar$, with
\begin{equation}\label{eq:choice-of-constant-for-power-constraint}
 c = \frac{1}{\sqrt{\trace(\bT_0 \bT_0^H)}}.
\end{equation}
\end{theorem}
\begin{IEEEproof}
Without loss of generality, we can express $\bD$ as
\begin{equation}\label{eq:diag-matrix}
 \bD = c \Dbar
\end{equation}
where $c>0$ and $\Dbar = \diag(\dbar_1,\ldots,\dbar_K)$ satisfies
\begin{equation}\label{eq:product-diagonal}
 |\det\Dbar|=1.
\end{equation}
Specifically, take $c = \left(\prod_{i=1}^K |d_i|\right)^{1/K}$ and $\dbar_i = d_i/c$, $i = 1,\ldots,K$.
This implies that
\begin{equation}\label{eq:proof-reifhi-intermediate}
\rdifhi(\bA,\bD) = \log_2\left(c^{2K} \SNR^K \right).
\end{equation}

Note also that this choice results in $\bT = c \bT_0$, with $\bT_0$ given in \eqref{eq:aux-precoding-matrix}. Now, since $1 \geq \trace(\bT \bT^H) = c^2\trace(\bT_0 \bT_0^H)$, it is easy to see that the rate is maximized under the power constraint by choosing $c$ according to \eqref{eq:choice-of-constant-for-power-constraint}. Replacing this value in \eqref{eq:proof-reifhi-intermediate} gives the desired result.
\end{IEEEproof}
\medskip

It follows from Theorem~\ref{thm:rate-hi-snr-objfun} that, for high SNR, optimizing the DIF scheme amounts to minimizing $\trace(\bT_0 \bT_0^H)$ under a constraint on $\Dbar$.

\subsection{Optimization in the Reverse Order}
\label{ssec:optimization-reverse}

We have seen that 
\begin{equation}
\rdifhi \triangleq \max_{\bA: \rank \bA = K}\; \max\limits_{\Dbar: |\det\Dbar|=1}\;  K \log_2\left(\frac{\SNR}{\trace(\bT_0 \bT_0^H)} \right)
\end{equation}
where $\bT_0 = \bH^H (\bH\bH^H)^{-1} \Dbar \bA$. Now suppose we reverse the order of maximization, i.e., we fix $\Dbar$ and wish to choose the optimal $\bA$. It can be shown that this problem can be converted into the shortest independent vector problem (SIVP) on a lattice \cite{Micciancio.Goldwasser.2002.Complexity-Lattice-Problems}.

Let $\bG_0 = \bH^H (\bH\bH^H)^{-1} \Dbar$, so that $\bT_0 = \bG_0 \bA$. We have
\begin{equation}
\trace(\bT_0^H \bT_0) = \sum_{i=1}^{K} \| \bT_0(i) \|^2 = \sum_{i=1}^K \|\bG_0 \bA(i)\|^2
\end{equation}
where $\bT_0(i)$ and $\bA(i)$ denote the $i$th \textit{column} of matrices $\bT_0$ and $\bA$, respectively. 

Since the columns of $\bA$ must be linearly independent, minimizing the above expression is equivalent to finding the $K$ shortest independent vectors of a lattice with generator matrix $\bG_0$, written in \textit{column} notation. Such vectors will produce the columns of $\bA$.

Note that the above method is a natural generalization of the method proposed in \cite{Hong2012} (see also \cite{Stern.Fischer.2016.Advanced-Factorization-Strategies}), which applies to the special case $\Dbar = \bI$.

In practice, the SIVP can be approximately solved by approximation algorithms for lattice basis reduction, such as the LLL algorithm \cite{Lenstra.etal.1982.Factoring-Polynomials-Rational,Gan.etal.2009.Complex-Lattice-Reduction}. However, due to the lack of an analytical expression for the optimal $\bA$, searching for the optimal $\Dbar$ (in conjunction with the above method) becomes a very difficult task. Indeed, the resulting objective function for $\Dbar$ is highly nonsmooth since it involves a discrete optimization step, making the overall problem prohibitively complex.

Even though we do not pursue this approach in this paper, we use it in Section~\ref{sec:Simulation Results} to demonstrate the potential of the proposed scheme for $K>2$, for which an analytical solution is still unknown. In the next section, we focus on first optimizing $\Dbar$ for a given $\bA$, as discussed before.

\section{The Two-User Case for High SNR}
\label{sec:two-user}

In this section, we restrict our attention to the special case of $K=2$ receivers in the high SNR regime, assuming that DIF is used.

\subsection{Optimal Signal Precoding}
\label{ssec:optimal-signal-precoding}

We start by optimizing the matrix $\Dbar$ in \eqref{eq:rate-hi-snr-objfun} and finding the resulting achievable rate.

Let
\begin{equation}\label{rho:exp}
\rho(\bH) \triangleq \frac{|\bh_1 \bh_2^H|}{\|\bh_1\|\|\bh_2\|}
\end{equation}
denote the normalized inner product between the rows of $\bH$,
and define
\begin{equation}\label{eq:obj-fun-coefficient-matrix}
f(\bA,\rho) \triangleq \|\ba_1\| \|\ba_{2}\| - \rho |\ba_2 \ba_1^H|.
\end{equation}

\medskip
\begin{theorem}\label{thm:rate-optimal-signal-precoding}
For any full-rank $\bA \in \ZZ[i]^{2 \times 2}$,
\begin{equation}\label{eq:rate-optimal-signal-precoding}
\rdifhi(\bA) =  
2 \log_2^+\left(\frac{ \det(\bH\bH^H) \cdot \SNR}{2\|\bh_{1}\| \|\bh_{2}\| \cdot  f(\bA,\rho(\bH))} \right)
\end{equation}
achievable with
\begin{equation}\label{eq:diag-matrix-opt}
 \Dbar = \mat{\sqrt{\frac{\|\ba_{2}\|\|\bh_1\|}{\|\ba_{1}\|\|\bh_2\|}}  & 0 \\ 0 & \sqrt{\frac{\|\ba_{1}\|\|\bh_2\|}{\|\ba_{2}\|\|\bh_1\|}}e^{-j\angle (\ba_2 \ba_1^H \bh_1 \bh_2^H)} }.
\end{equation}
\end{theorem}

\begin{IEEEproof}
Recall that $\Dbar = \diag(\dbar_1,\dbar_2)$.
Without loss of generality, let $\dbar_1 = e^{\beta+j\theta_1}$ and $\dbar_2 = e^{-\beta+j\theta_2}$, where \mbox{$\beta \in \RR$}. Note that $\det(\Dbar) = |\dbar_1| |\dbar_2| = 1$. 
In order to simplify notation, let
\begin{equation}
 \bM = (\bH\bH^H)^{-1} = \frac{1}{\det(\bH\bH^H)}\mat{\|\bh_2\|^2 & -\bh_1 \bh_2^H \\  -\bh_2 \bh_1^H & \|\bh_1\|^2}
\end{equation}
with entries $M_{ij}$. Since $(\bA \bA^H)_{ij} = \ba_i \ba_j^H$,
\begin{equation}
 (\Dbar^H\bM\Dbar)_{ij} = \dbar_i^* M_{ij} \dbar_j
\end{equation}
and $\bM^H = \bM$, we have that
\begin{align}
\trace(\bT_0^H \bT_0) 
&= \trace\left( \bA^H \Dbar^H \bM \Dbar \bA \right)\nonumber\\
&= \trace\left( \bA \bA^H \Dbar^H \bM \Dbar \right) \nonumber\\
&= \|\ba_1\|^2 M_{11} e^{2\beta} + \|\ba_2\|^2 M_{22} e^{-2\beta} \nonumber \\
&\hphantom{=.} + 2\Re\{\ba_2 \ba_1^H M_{12} e^{j\Delta \theta}\} \label{eq:signal-precoding-obj-function}
\end{align}
where $\Delta \theta = \theta_2 - \theta_1$.

We wish to minimize \eqref{eq:signal-precoding-obj-function} by the choice of $\beta$ and $\Delta \theta$.
Clearly, the optimal choice of $\Delta \theta$ is
\begin{equation}
 \Delta\theta = - \angle (- \ba_2 \ba_1^H M_{12}) = -\angle ( \ba_2 \ba_1^H \bh_1 \bh_2^H)
\end{equation}
which gives
\begin{equation}
\trace(\bT_0^H \bT_0) 
= \|\ba_1\|^2 M_{11} e^{2\beta} + \|\ba_2\|^2 M_{22} e^{-2\beta} - 2|\ba_2 \ba_1^H M_{12}|.
\end{equation}

Solving for the optimal $\beta$, we have
\begin{align}
0 
&= 
\textstyle\frac{\partial}{\partial\beta} \trace(\bT_0^H \bT_0) \nonumber \\
&= 2 \|\ba_{1}\|^2 M_{11} e^{2\beta} - 2 \|\ba_{2}\|^2 M_{22} e^{-2\beta}
\end{align}
which gives
\begin{equation}
 e^{2\beta} = \frac{\|\ba_{2}\|\sqrt{M_{22}}}{\|\ba_{1}\|\sqrt{M_{11}}} = \frac{\|\ba_{2}\|\|\bh_1\|}{\|\ba_{1}\|\|\bh_2\|}.
\end{equation}

Substituting this optimal choice of $\beta$, we have
\begin{align}
\trace(\bT_0^H \bT_0)
&= 2 \|\ba_1\| \|\ba_{2}\| \sqrt{M_{11}M_{22}} - 2|\ba_2 \ba_1^H M_{12}| \nonumber \\
&= 2 \sqrt{M_{11}M_{22}}(\|\ba_1\| \|\ba_{2}\| - \rho |\ba_2 \ba_1^H|) \nonumber \\
&= 2 \sqrt{M_{11}M_{22}} f(\bA,\rho) \nonumber \\
&= 2 \frac{\|\bh_1\| \|\bh_2\|}{\det(\bH \bH^H)} f(\bA,\rho) \label{eq:trace-final}
\end{align}
where
\begin{equation}
 \rho = \rho(\bH) = \frac{|M_{12}|}{\sqrt{M_{11}M_{22}}}.
\end{equation}

The result now follows by replacing \eqref{eq:trace-final} in \eqref{eq:rate-hi-snr-objfun} and setting $\theta_1 = 0$.
\end{IEEEproof}
\medskip

It follows from Theorem~\ref{thm:rate-optimal-signal-precoding} that
\begin{equation}\label{eq:rate-optimal}
\rdifhi =  
2 \log_2\left(\frac{ \det(\bH\bH^H) \cdot \SNR}{2\|\bh_{1}\| \|\bh_{2}\| \cdot  f(\rho(\bH))} \right)
\end{equation}
where
\begin{equation}\label{eq:optim-problem-coefficient-matrix}
f(\rho) \triangleq \min_{\bA \in \ZZ[j]^{2 \times 2}: \rank \bA = 2}\, f(\bA,\rho).
\end{equation}

\subsection{Optimal Message Precoding}
\label{ssec:optimal-message-precoding}

We now address the optimal choice of the message precoding matrix $\bA$ in \eqref{eq:optim-problem-coefficient-matrix}.

First, we need a few definitions. 
Let
\begin{equation}
 \calN_2 = \{|a|^2: a \in \ZZ[j]\} = \{a^2 + b^2: a,b \in \ZZ\} \subseteq \ZZ
\end{equation}
be the set of all integers that are sums of two squares. For any $x \in \RR$, let $\lfloor x \rfloor_{\calN_2}$ denote the largest element of $\calN_2$ smaller than or equal to $x$. Similarly, let $\lceil x \rceil_{\calN_2}$ denote the smallest element of $\calN_2$ greater than or equal to $x$.
\medskip
\begin{theorem}
We have
\begin{equation}\label{eq:optim-problem-by-norm}
f(\rho) = \min_{N \in \calN_2}\, \sqrt{N+1} - \rho \sqrt{N}.
\end{equation}
This value is
achievable by any full-rank $\bA$ satisfying
\begin{equation*}
\|\ba_1\|^2 \|\ba_{2}\|^2 -1 = |\ba_1 \ba_2^H|^2  = N
\end{equation*}
in particular, by
\begin{equation}\label{eq:optimal-solution-integer-matrix}
 \bA = \mat{\ba_1 \\ \ba_2} = \mat{1 & 0 \\ a_{21} & 1}
\end{equation}
where $a_{21} \in \ZZ[j]$ is such that $|a_{21}|^2 = N$.
\end{theorem}

\begin{IEEEproof}
Recall that $f(\bA,\rho) = \|\ba_1\| \|\ba_{2}\| - \rho |\ba_2 \ba_1^H|$.
Our approach to solve \eqref{eq:optim-problem-coefficient-matrix} is to find
a pair of linearly independent vectors $\ba_1,\ba_2 \in \ZZ[j]^2$ that minimizes $\|\ba_1\| \|\ba_{2}\|$, for every possible value of $|\ba_1 \ba_2^H|^2 \in \calN_2$. 

For all $N \in \calN_2$, let
\begin{equation}\label{eq:minimization-prod-norm-constrained}
 \pi_N \triangleq 
 \min_{\ba_1,\ba_2: |\ba_1 \ba_2^H|^2 = N}\; \|\ba_1\| \|\ba_{2}\|
\end{equation}
where the minimization is restricted to $\ba_1,\ba_2 \in \ZZ[j]^2$ that are linearly independent. It follows that $f(\rho)$ is equal to the minimium value of $\pi_N - \rho \sqrt{N}$ over all $N \in \calN_2$.

From the Cauchy-Schwarz inequality, we know that $\|\ba_1\|^2 \|\ba_{2}\|^2 \geq |\ba_1 \ba_2^H|^2$, achievable if and only if $\ba_2$ is a multiple of $\ba_1$. However, this condition violates the requirement of linear independence. Thus, since both $\|\ba_1\|^2 \|\ba_{2}\|^2$ and 
$|\ba_1 \ba_2^H|^2$ must be integers, we must have
\begin{equation}\label{eq:optimal-constraint-coefficient-vectors}
\|\ba_1\|^2 \|\ba_{2}\|^2 \geq |\ba_1 \ba_2^H|^2 + 1.
\end{equation}

Since equality is always achievable, e.g., by \eqref{eq:optimal-solution-integer-matrix},
we have that $\pi_N^2 = N+1$, for all $N \in \calN_2$.
\end{IEEEproof}
\medskip

Table~\ref{glinha} lists all possible \textit{non-equivalent} solutions of \eqref{eq:optim-problem-coefficient-matrix} for $N \leq~20$.
Equivalent solutions can be found by permuting rows and/or columns and by multiplying rows and/or columns by $-1$, $j$, or $-j$.
Note that there can be multiple non-equivalent solutions for certain values of $N$.

\begin{table}[t]
\center
\caption{Non-equivalent solutions to optimal integer matrix $\bA$ and respective $\rho_N$ values for $N \leq 20$.}
\begin{tabular}{ccccc}
\toprule
$\ba_1$ & $\ba_2$ & $\|\ba_1\|^2\|\ba_{2}\|^2$ & $N = |\ba_1 \ba_2^H|^2$ & $\rho_N $ \\ 
\midrule
$(1,0)$ & $(0,1)$ & 1 & 0 & 0 \\ 
$(1,0)$ & $(1,1)$ & 2 & 1 & 0.4142 \\
$(1,0)$ & $(1+j,1)$ & 3 & 2 & 0.7673 \\
$(1,0)$ & $(2,1)$ & 5 & 4 & 0.8604 \\
$(1,0)$ & $(2+j,1)$ & 6 & 5 & 0.9041 \\
$(1,1)$ & $(1+j,1)$ &  &  & \\ 
$(1,0)$ & $(2+2j,1)$ & 9 & 8 & 0.9294 \\
$(1,0)$ & $(3,1)$ & 10 & 9 & 0.9458 \\
$(1,1)$ & $(2,1)$ &  &  & \\ 
$(1,0)$ & $(3+j,1)$ & 11 & 10 & 0.9511 \\
$(1,0)$ & $(3+2j,1)$ & 14 & 13 & 0.9588\\
$(1,0)$ & $(4,1)$ & 17 & 16 & 0.9670\\
$(1,0)$ & $(4+j,1)$ & 18 & 17 & 0.9710 \\
$(1+j,1)$ & $(2+j,1)$ &  &  & \\ 
$(1,0)$ & $(3+3j,1)$ & 19 & 18 & 0.9726\\
$(1,0)$ & $(4+2j,1)$ & 21 & 20 & 0.9746\\
\bottomrule
\end{tabular} 
\label{glinha}
\end{table}

The next theorem shows that the optimal value of $N$ can be found in an almost closed form.

\medskip
\begin{theorem}\label{thm:optimal-value-norm}
The optimal value of $N$ in \eqref{eq:optim-problem-by-norm} satisfies
\begin{equation}\label{eq:optim-value-norm}
 N \in \left\{ \left\lfloor \frac{\rho^2}{1-\rho^2}\right\rfloor_{\calN_2}, \left\lceil \frac{\rho^2}{1-\rho^2}\right\rceil_{\calN_2} \right\}.
\end{equation}
Moreover, each value of $N \in \calN_2$ is an optimal solution for $\rho_N \leq \rho \leq \rho_{N^+}$, where $\rho_0 \triangleq 0$,
\begin{equation}
\rho_N \triangleq \frac{\sqrt{N+1}-\sqrt{N^-+1}}{\sqrt{N}-\sqrt{N^-}}, \quad N \geq 1,
\end{equation}
$N^- = \lfloor N-1 \rfloor_{\calN_2}$ and $N^+ = \lceil N+1 \rceil_{\calN_2}$.
\end{theorem}
\begin{IEEEproof}
Let 
\begin{equation}
 f(N,\rho) = \sqrt{N+1} - \rho \sqrt{N}.
\end{equation}
We have that $f(N,\rho) \leq f(N^-,\rho)$ if and only if
\begin{equation}
 \sqrt{N+1} - \rho\sqrt{N} \leq \sqrt{N^-+1} - \rho\sqrt{N^-}
\end{equation}
i.e., if and only if $\rho \geq \rho_N$.
It follows that $N$ is optimal for all~$\rho$ satisfying $\rho_N \leq \rho \leq \rho_{N^+}$.

Now, suppose $\rho$ is fixed and consider the relaxed function
$g(x) = \sqrt{x+1}- \rho \sqrt{x}$, where $x \in \RR$.
This function
has a single critical point which is a global minimum, given by
\begin{equation}
 x^* = \frac{\rho^2}{1-\rho^2}.
\end{equation}
It follows that the optimal value of $N$ is either the floor or ceiling of $x^*$ in $\calN_2$, i.e., $N$ must satisfy \eqref{eq:optim-value-norm}.
\end{IEEEproof}
\medskip

As a consequence of Theorem~\ref{thm:optimal-value-norm}, we conclude that finding the optimal $N$, and thus the optimal $\bA$, amounts simply to nonuniform scalar quantization based on table look-up, so its complexity is very low.

It is worth treating the special case where $\bA$ is constrained to be a real integer matrix, $\bA \in \ZZ^{2 \times 2}$, in which case $N$ must be of the form $N = k^2$, where $k \in \ZZ$. Although suboptimal, this choice gives an upper bound on $f(\rho)$ which is easier to analyze.

\medskip
\begin{theorem}\label{thm:optimal-value-real}
We have
\begin{equation*}
f(\rho) \leq \min_{k \in \ZZ,\, k \geq 0}\, \sqrt{k^2+1} - \rho k
\end{equation*}
achievable by some $k \in \ZZ$ satisfying
\begin{equation}
 k \in \left\{ \left\lfloor u \right\rfloor, \left\lceil u \right\rceil \right\}
\end{equation}
where $u = \rho/\sqrt{1-\rho^2}$. Moreover, each value of $k \geq 0$ is an optimal solution for $u_k \leq u \leq u_{k+1}$, where $u_0 = 0$ and, for $k\geq 1$, $u_k$ is defined by
\begin{equation}\label{eq:transition-values-real}
\frac{1}{\sqrt{1/u_k^2+1}} = \sqrt{k^2+1}-\sqrt{(k-1)^2+1}.
\end{equation}
\end{theorem}
\begin{IEEEproof}
Except for a change of variables, the proof is very similar to that of Theorem~\ref{thm:optimal-value-norm} and is therefore omitted.
\end{IEEEproof}
\medskip

\begin{corollary}
For all $k \geq 0$, $\lceil u_k \rceil = \lfloor u_{k+1} \rfloor = k$.
\end{corollary}
\begin{IEEEproof}
The statement is true for $k=0$, as can be easily checked. For $k\geq 1$, assume the statement is true for $k-1$, i.e., $\lceil u_{k-1} \rceil = \lfloor u_{k} \rfloor = k-1$. From Theorem~\ref{thm:optimal-value-real}, we know that
\begin{equation*}
k \in \left\{ \lfloor u_k \rfloor, \lceil u_k \rceil \right\} \cap \left\{ \lfloor u_{k+1} \rfloor, \lceil u_{k+1} \rceil \right\}.
\end{equation*}
Since $u_k \not\in \ZZ$ and $u_k < u_{k+1}$, this implies that either
\begin{equation}\label{eq:proof-hypothesis-1}
\lfloor u_k \rfloor = \lfloor u_{k+1} \rfloor < \lceil u_k \rceil = \lceil u_{k+1} \rceil
\end{equation}
or
\begin{equation}\label{eq:proof-hypothesis-2}
\lceil u_k \rceil = \lfloor u_{k+1} \rfloor = k.
\end{equation}
By the induction hypothesis, \eqref{eq:proof-hypothesis-1} implies that $\lfloor u_{k+1}\rfloor = k-1$ and $\lceil u_{k+1} \rceil = k$, which is a contradiction since, from Theorem~\ref{thm:optimal-value-real}, we must have $k+1 \in \left\{ \lfloor u_{k+1} \rfloor, \lceil u_{k+1} \rceil \right\}$. Thus, \eqref{eq:proof-hypothesis-2} must be true, proving the statement.
\end{IEEEproof}

\subsection{Gap to Sum Capacity}
\label{ssec:gap}

We now investigate the asymptotic gap of DIF to the sum capacity, $\csum$, given in (\ref{eq:sum:cap}). 

\medskip
\begin{theorem}\label{thm:gap-to-capacity}
Let $\rho = \rho(\bH)$. We have
\begin{align*}
\lim_{\SNR \to \infty} \csum - \rdifhi 
&= 2\log_2 \left(\frac{f(\rho)}{\sqrt{1-\rho^2}}\right) \\
&\leq \log_2\left( \frac{1+\sqrt{2}}{2} \right).
\end{align*}
\end{theorem}

\begin{IEEEproof}
It is known from \cite{Lee2007} that, for high SNR, 
\begin{equation}
 \lim_{\SNR \to \infty} \csum - \csumhi = 0
\end{equation}
where
\begin{equation}\label{HighSNRSUMCap}
\csumhi = K \log_2 (\SNR/K) + \log_2 \det(\bH \bH^H)
\end{equation}
with $K=2$ in the present case.

Let $\delta = \csumhi - \rdifhi$ and $\Delta = 2^{\delta/2}$. We have that
\begin{align*}
\Delta 
&= \frac{\frac{1}{2}\SNR \sqrt{\det(\bH\bH^H)}}{\left(\frac{ \det(\bH\bH^H)\cdot \SNR}{2\|\bh_{1}\| \|\bh_{2}\|}\cdot \frac{1}{ f(\rho)} \right)}\\
&= \frac{\|\bh_{1}\| \|\bh_{2}\|}{\sqrt{\det(\bH\bH^H)}} f(\rho) \\
&= \frac{f(\rho)}{\sqrt{1-\rho^2}}
\end{align*}
where the last equality follows since
\begin{equation*}
\frac{\det(\bH\bH^H)}{\|\bh_{1}\|^2 \|\bh_{2}\|^2} = 1 - \rho(\bH)^2.
\end{equation*}

We now proceed to give an upper bound on $\Delta$. First, note that, from Theorem~\ref{thm:optimal-value-real},
\begin{equation*}
\Delta \leq \Delta_{k^*}(u)
\end{equation*}
where $u = \rho/\sqrt{1-\rho^2}$,
\begin{equation*}
\Delta_k(u) = \sqrt{u^2+1}\sqrt{k^2+1} - u k
\end{equation*}
and $k^* \geq 0$ is such that $u_{k^*} \leq u \leq u_{k^*+1}$.

Now, $\Delta_k(u)$ is a convex function of $u$, which implies that
\begin{align*}
\Delta_{k^*}(u) 
&\leq \max\{ \Delta_{k^*}(u_{k^*}),\, \Delta_{k^*}(u_{k^*+1}) \} \\
&\leq \max_{k \geq 0}\, \Delta_k(u_k)
\end{align*}
where in the last equation we have used the fact that $\Delta_k(u_k) = \Delta_k(u_{k+1})$, for all $k$.

Numerical evaluation reveals that $\Delta_0(u_0) = 1$,
\begin{equation*}
\Delta_1(u_1) = \sqrt{\frac{1+\sqrt{2}}{2}}
\end{equation*}
and $\Delta_1(u_1) \geq \Delta_2(u_2) \geq \Delta_3(u_3)$. For larger $k$, note that
\begin{align*}
\Delta_k(u_k) 
&= \sqrt{u_k^2k^2+u_k^2 + k^2 + 1} - u_k k \\
&= u_kk\sqrt{1 + \frac{u_k^2 + k^2 + 1}{u_k^2k^2}} - u_k k \\
&\leq u_kk\left(1 + \frac{u_k^2 + k^2 + 1}{2u_k^2k^2}\right) - u_k k \\
&= \frac{u_k^2 + k^2 + 1}{2u_kk} \\
&= \frac{(k-u_k)^2 + 2u_kk + 1}{2u_kk} \\
&= 1 + \frac{(k-u_k)^2 + 1}{2u_kk} \\
&\leq 1 + \frac{1}{k^2}
\end{align*}
where the last inequality follows since $0 \leq k-u_k < 1$. It can be easily checked that $\Delta_k(u_k) \leq \Delta_1(u_1)$ for $k \geq 4$, completing the proof.
\end{IEEEproof}
\medskip

Theorem~\ref{thm:gap-to-capacity} shows that, for high SNR, DIF achieves a small gap to sum capacity, upper bounded by about $0.27$~bits, or approximately $0.4$~dB. This maximum value occurs for $\rho = \sqrt{2}-1$ in the transition from $N=0$ to $N=1$ in Table~\ref{glinha}. 

\begin{figure}
\centering
\includegraphics[scale=0.66]{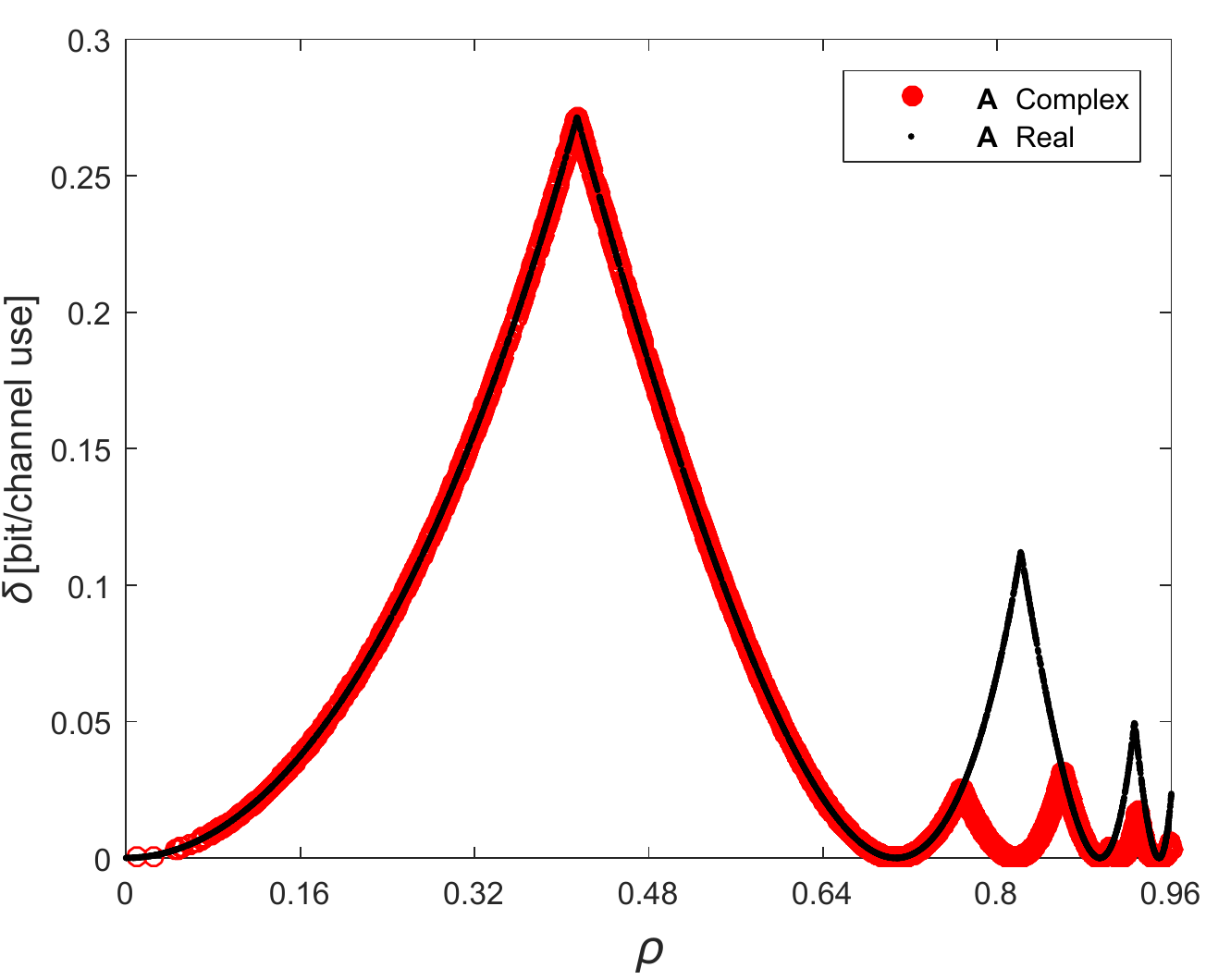}
\caption{Gap from the DIF achievable rate to the sum capacity in the high SNR regime.}
\label{Gap}
\end{figure}

The gap to sum capacity as a function of $\rho$ is illustrated in Fig.~\ref{Gap} for both the complex case $\bA\in \ZZ[i]^{2 \times 2}$ (Theorem~\ref{thm:optimal-value-norm}) and the real case $\bA\in \ZZ^{2 \times 2}$ (Theorem~\ref{thm:optimal-value-real}). 
Peaks occur at values $\rho = \rho_N$ given in Table~\ref{glinha} for the complex case, and at $\rho$ equal to the right hand side of \eqref{eq:transition-values-real} for the real case.
Interestingly, the gap vanishes both as $\rho \to 0$ (orthogonal channels) and as $\rho \to 1$.

\section{Extension to General SNR}
\label{sec:General SNR}

We have shown in Section~\ref{sec:exact-if} that, among all integer-forcing precoding schemes, DIF is optimal for high SNR. For general SNR, however, designing an optimal integer-forcing precoder is much more challenging, since we have much more freedom in the choice of the precoding matrix~$\bT$, besides the choice of~$\bA$. 
As noted before, even the special case of $\bA = \bI$ , which is the problem of optimal linear precoding, is already NP-hard.

As in the case of conventional linear precoding, one approach to find a good suboptimal solution may be to propose a heuristic structure for the precoder and optimize it with respect to either the original cost function or some more tractable approximation.

\subsection{Regularized DIF}

For finite SNR, we propose to use the following precoding structure:
\begin{align}
\bT &= c \bT_0 \\
\bT_0 &= \bH^H \left(\frac{K}{\SNR}\bI + \bH\bH^H\right)^{-1} \Dbar \bA \label{eq:aux-precoding-matrix-regularized}
\end{align}
where $c > 0$ is chosen as in \eqref{eq:choice-of-constant-for-power-constraint} in order to satisfy the power constraint. We call this method \textit{regularized} DIF (RDIF). This structure (and our choice of terminology) is inspired by regularized~ZF (RZF) precoding, which corresponds to the special case $\bA = \bI$, in the same way that ZF is a special case of DIF. Note also that, as SNR grows, \eqref{eq:aux-precoding-matrix-regularized} tends to \eqref{eq:aux-precoding-matrix}, and therefore RDIF reduces to DIF, as we should expect. A formal justification for the RDIF structure is provided in the next subsection.

Given the RDIF structure, consider now the problem of optimally choosing $\Dbar$ and $\bA$. In this case, since we do not have a simple expression for the achievable sum rate \eqref{eq:sum-rate-optimized-precoding}, instead we choose to maximize the high SNR lower bound \eqref{eq:rate-hi-snr-objfun}, which is equivalent to minimizing $\trace(\bT_0^H \bT_0)$.

It follows that we can easily adapt all the results in sections~\ref{ssec:optimization-reverse}, \ref{ssec:optimal-signal-precoding} and \ref{ssec:optimal-message-precoding} by replacing $\bM = (\bH\bH^H)^{-1}$ with $\bM = \left(\frac{K}{\SNR}\bI + \bH\bH^H\right)^{-1}$. In particular, for $K=2$, an optimal (in the sense of maximizing \eqref{eq:rate-hi-snr-objfun}) choice of $\Dbar$ is given by
\begin{equation}
\Dbar = \mat{\sqrt{\frac{\|\ba_{2}\|\sqrt{M_{22}}}{\|\ba_{1}\|\sqrt{M_{11}}}} & 0 \\ 0 & \sqrt{\frac{\|\ba_{1}\|\sqrt{M_{11}}}{\|\ba_{2}\|\sqrt{M_{22}}}} e^{-j\angle (- \ba_2 \ba_1^H M_{12})}}
\end{equation}
while that of $\bA$ can be found by applying the method of Section~\ref{ssec:optimal-message-precoding} with the value of $\rho$ given by
\begin{equation}
 \rho = \rho(\bH) = \frac{|M_{12}|}{\sqrt{M_{11}M_{22}}}.
\end{equation}
In this case, since the method of Section~\ref{ssec:optimal-message-precoding} can be easily implemented by table look-up, the overall complexity is essentially that of $2\times 2$-matrix inversion and multiplication, similar to that of RZF.

For $K>2$, however, finding an optimal $\Dbar$ from a given $\bA$, as well as an optimal $\bA$, is still an open problem.

\subsection{Justification via Uplink-Downlink Duality}

We can give a formal justification for RDIF based on the uplink-downlink duality for integer forcing proved in \cite{He2014}.
For brevity, we leave out the details of the work in \cite{He2014} and only sketch the main ideas here.

It is shown in \cite{He2014} that any rate tuple achievable in the downlink with a certain power vector can also be achieved in the uplink with a certain (possibly different) power vector, and the same result holds with the roles of downlink and uplink reversed. Each power coefficient refers to the second moment of the shaping lattice for the corresponding user (in the case where multiple shaping lattices are allowed). In addition, the uplink and downlink channels are subject to the same total power constraint and are related by a transpose (or Hermitian, in the complex case) of all corresponding matrices, similarly to the general uplink-downlink duality for the MAC and the BC.

By starting with the downlink problem with a rate tuple that achieves a certain sum rate $R$, we can create a virtual dual uplink problem where the same rate tuple is achievable. For that problem, with all other parameters fixed, the optimal beamforming matrix can be obtained in closed form, which can only possibly improve the sum rate. Next, going back to original downlink problem, we have a solution that achieves a sum rate equal or higher than $R$. This is the basis for an iterative optimization algorithm proposed in \cite{He.etal.2014.Uplink-downlink-Duality-Integer-forcing:Iterative-Optimization}.
Here, however, we note that if the downlink beamforming matrix were already in a form that would produce an optimal uplink beamforming matrix, then this iteration of the algorithm would provide no improvement to the sum rate, i.e., such a matrix would be a fixed point of the iterative algorithm in \cite{He.etal.2014.Uplink-downlink-Duality-Integer-forcing:Iterative-Optimization}. 

Thus, as a consequence of \cite[(27)--(29)]{He.etal.2014.Uplink-downlink-Duality-Integer-forcing:Iterative-Optimization}, we can claim that, for any given $\bH$, $\bA$ and $\SNR$, at least one fixed point of the algorithm in \cite{He.etal.2014.Uplink-downlink-Duality-Integer-forcing:Iterative-Optimization} can be achieved with a beamforming matrix of the form\footnote{In the notation of \cite{He.etal.2014.Uplink-downlink-Duality-Integer-forcing:Iterative-Optimization}, $\bD = \bC_d^{-1}$, $\bQ = \bC_u \bP_{c,u} \bC_u^H$ and $\bC_d = \bC_u^H$, all of which are diagonal matrices.}
\begin{align}
\bT 
&= (\bI + \bH^H \bQ \bH)^{-1} \bH^H \bQ \bD \bA \\
&= \bH^H (\bI + \bQ \bH \bH^H)^{-1} \bQ \bD \bA \\
&= \bH^H (\bQ^{-1} + \bH \bH^H)^{-1} \bD \bA \label{eq:optimal-general-beamforming-matrix}
\end{align}
for some diagonal matrix $\bD \in \CC^{K \times K}$ and some nonnegative diagonal matrix $\bQ \in \RR^{K \times K}$ satisfying $\trace(\bQ) = \SNR$ and $\trace(\bT^H \bT) = 1$. More precisely, such a fixed point is obtained by choosing $\bQ$ and $\bD$ that maximize the sum rate. The uplink iteration cannot improve the sum rate any further since any optimal uplink beamforming matrix is already in the form \eqref{eq:optimal-general-beamforming-matrix}.

It is worth mentioning that the iterative algorithm in \cite{He.etal.2014.Uplink-downlink-Duality-Integer-forcing:Iterative-Optimization} cannot be directly used in our problem, since there is no guarantee that the resulting downlink power vector will be symmetric, which is a main assumption in this paper. Of course, this issue is avoided when we already start with a symmetric downlink power vector at a fixed point.

The RDIF structure differs from the optimal in \eqref{eq:optimal-general-beamforming-matrix} by our heuristic choice of an all-equal diagonal matrix $\bQ = q \bI$. Since we must have $\trace(\bQ) = \SNR$, this immediately gives $q = \SNR/K$, simplifying the optimization problem. Note that our choice of $\bQ$ is analogous to the heuristic choice of all-equal Lagrange multipliers when RZF is derived from the optimal solution structure for conventional linear precoding \cite{Bjornson2014}.

\section{Numerical Results}
\label{sec:Simulation Results}

In this section, numerical results on the sum-rate performance of the proposed schemes are presented. These results correspond to an average over a total of 1000 channel realizations for each value of SNR. Independent Rayleigh fading is assumed, where $\bH$ has i.i.d entries $\sim \calC\calN(0,1)$.

\begin{figure}
\centering
\includegraphics[scale=0.67]{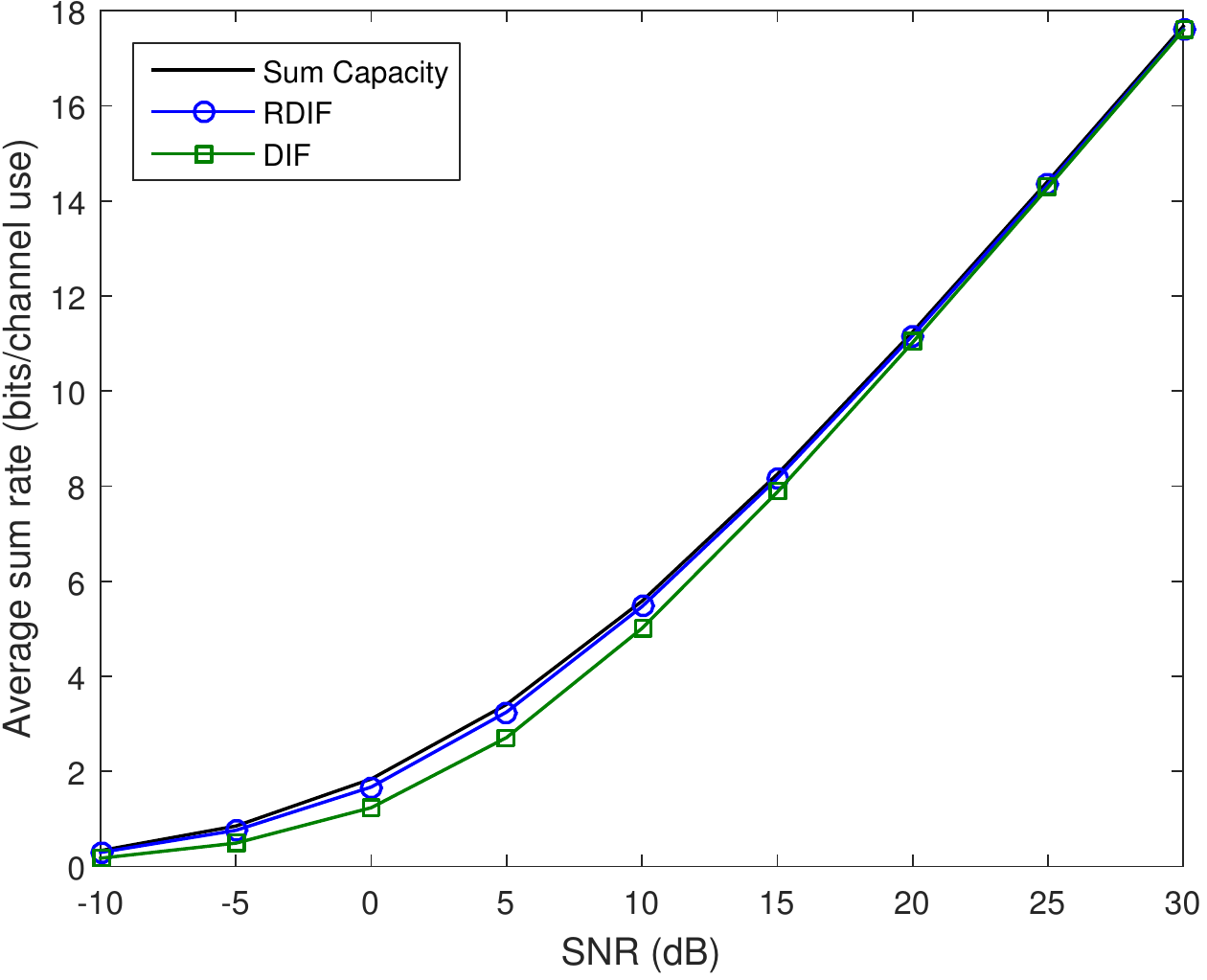}
\caption{Average sum rate for Rayleigh fading with $M = 2$ transmit antennas and $K = 2$ single-antenna receivers.}
\label{SumRate}
\end{figure}

We start with the case $M = K = 2$.
Fig.~\ref{SumRate} shows the average sum rate of DIF and RDIF over a large range of SNR values. For comparison, the sum capacity \eqref{eq:sum:cap} obtained by DPC \cite{Caire2003} is also shown. It can be seen that RDIF improves on the performance of DIF for low/medium SNR, while both schemes tend to the optimal performance for high SNR.

\begin{figure}
\centering
\includegraphics[scale=0.67]{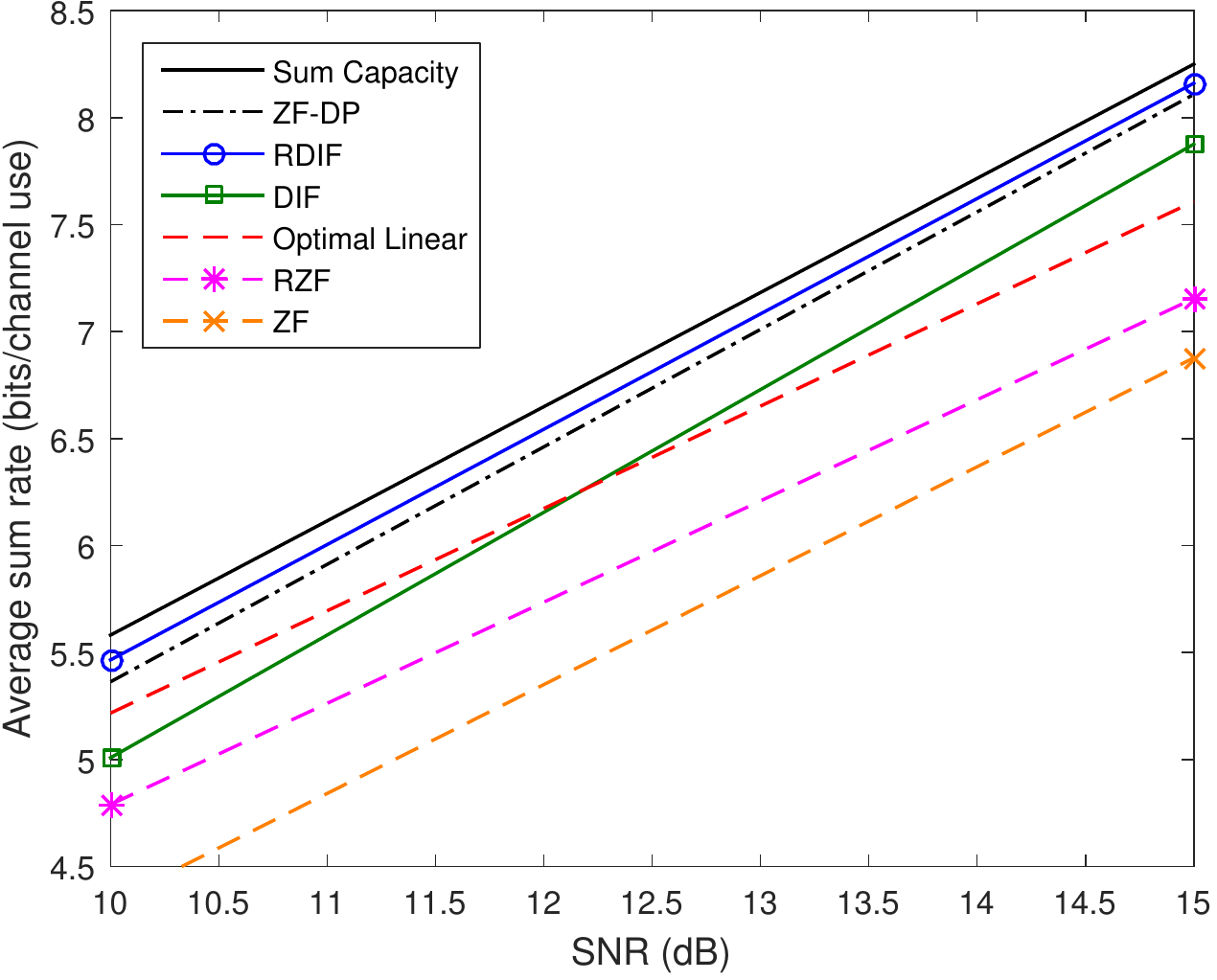}
\caption{Average sum rate of Fig.~\ref{SumRate} for the medium SNR region.}
\label{SumRate_db}
\end{figure}

Fig.~\ref{SumRate_db} shows a close up of Fig.~\ref{SumRate} on a range of medium SNR values. For comparison, the performances of several other precoders are also included, namely: the well-known linear precoders ZF and RZF, the optimal linear solution given in \cite{Bjornson2014}, and the nonlinear zero-forcing DPC (ZF-DP) \cite{Caire2003}. 
As can be seen, for an average sum rate of 6 bits/channel use, RDIF is less than 0.21~dB away from the sum capacity. 

\begin{figure}
\centering
\includegraphics[scale=0.67]{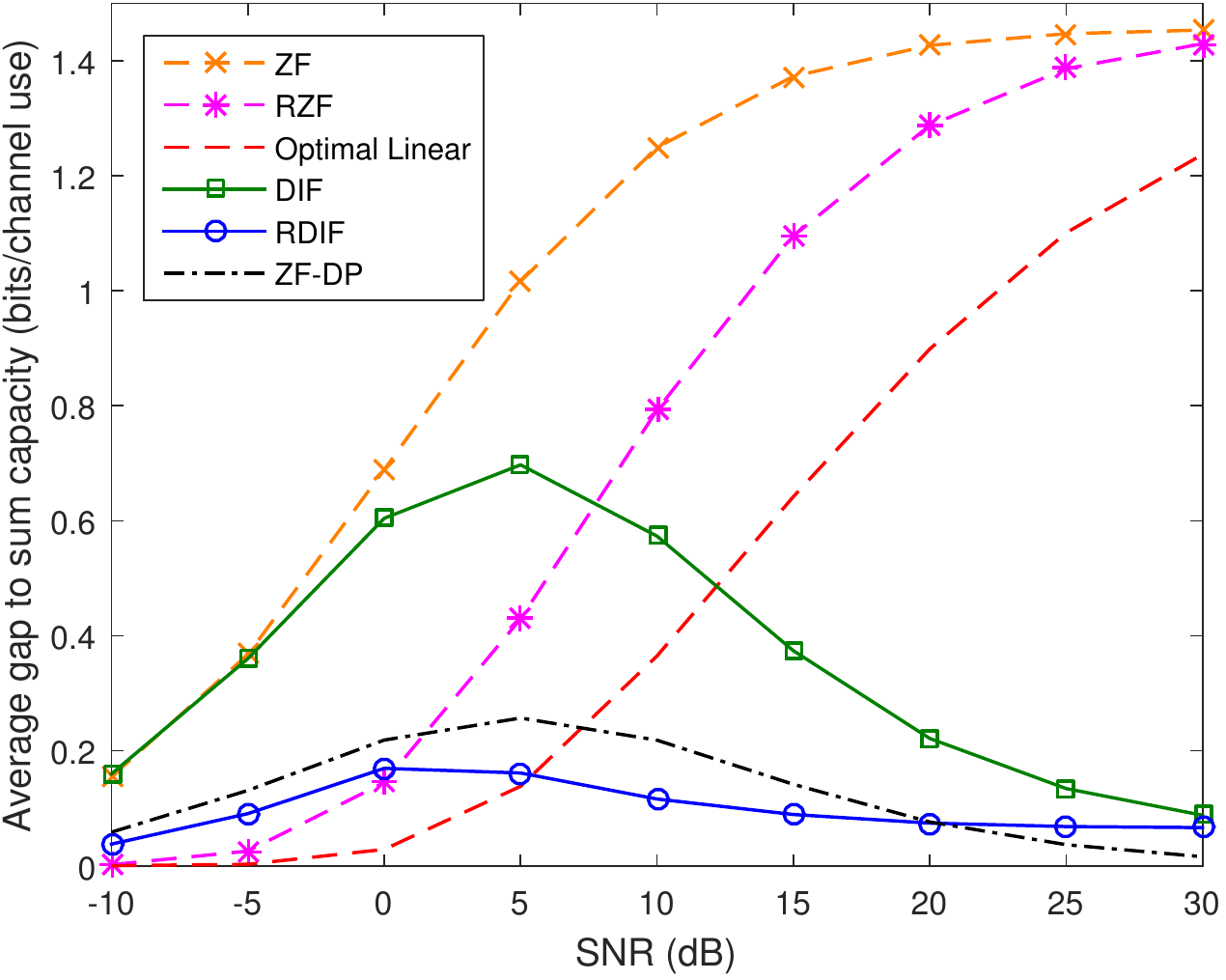}
\caption{Average gap to sum capacity for Rayleigh fading with $M = 2$ transmit antennas and $K = 2$ single-antenna receivers.}
\label{AverageGap}
\end{figure}

In order to clearly quantify the loss in performance compared to the optimal solution, Fig.~\ref{AverageGap} shows the average gap to sum capacity of all suboptimal schemes. As can be seen, for SNR $> 5.5$ dB, RDIF outperforms all conventional linear schemes. Moreover, RDIF outperforms ZF-DP for SNR $< 20$~dB. 
Both DIF and RDIF (as well as ZF-DP) have its worst performance for low to medium SNR. As predicted in Section~\ref{ssec:gap}, we can see that the two proposed schemes obey the gap of $\delta<0.27$ bits for high SNR, although the actual gap is much smaller on average. 
Note that, for low SNR, RDIF underperforms RZF, despite the fact that RZF is a special case of RDIF. This is because we design $\bA$ and $\Dbar$ based on the high SNR lower bound. At low SNR, the norms of the rows of $\bA$ have a significant impact on the sum rate, so ideally the method should instead apply some kind of penalty to such high-norm cases.

\begin{figure}
\centering
\includegraphics[scale=0.67]{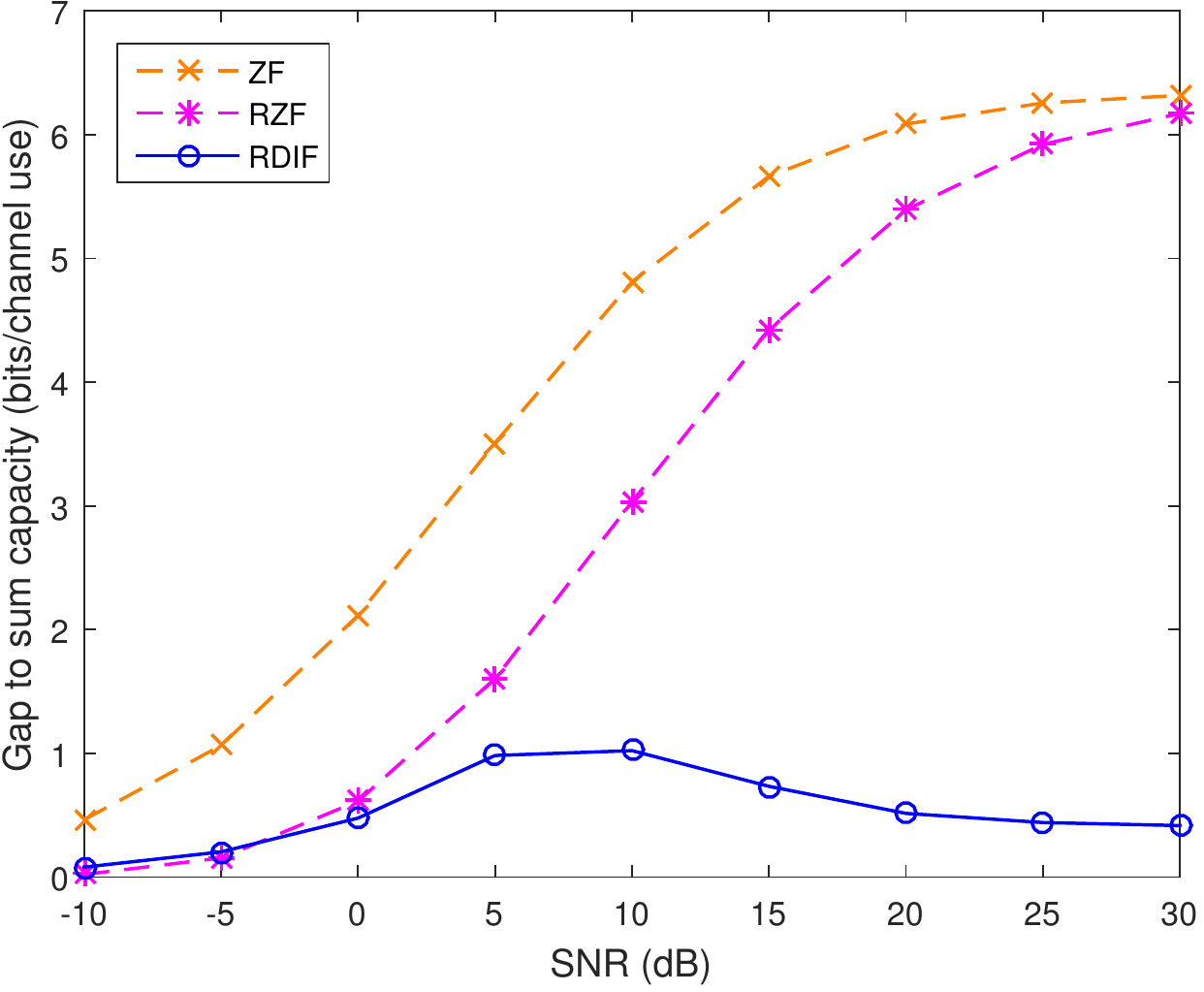}
\caption{Average gap to sum capacity for Rayleigh fading with $M = 4$ transmit antennas and $K = 4$ single-antenna receivers.}
\label{AverageGap4}
\end{figure}

While we currently only have an efficient design method for DIF/RDIF for $K=2$, it is possible to demonstrate the theoretical potential of these schemes for $K>2$ by globally searching for an optimal $\Dbar$ according to the method of Section~\ref{ssec:optimization-reverse}. Following this approach, Fig.~\ref{AverageGap4} shows the average gap to sum capacity for RDIF for the case $K=M=4$. For comparison, we also plot the performance of the RZF and ZF precoders. As can be seen, the performance of RDIF is also reasonably close to the sum capacity for all SNR and significantly outperforms ZF and RZF for high SNR.

\section{Conclusion}\label{sec:conclusion}

In this paper, we propose two integer-forcing precoding schemes for the Gaussian MIMO BC, called diagonally-scaled exact integer-forcing (DIF) and regularized DIF (RDIF). These precoders generalize ZF and RZF, respectively. Essentially, DIF creates an effective channel matrix that is exactly integer up to diagonal scaling, while RDIF modifies DIF with a certain matrix regularization in order to improve its performance for low to medium SNR.

We show that the DIF structure is optimal for high SNR, in the sense that it achieves maximum spatial multiplexing gain. Moreover, in the special case of two receivers and high SNR, we have shown that the optimal choice of parameters can be solved analytically.
In this case, the gap to sum capacity of the Gaussian MIMO BC is found to be upper bounded by 0.27 bits. For general SNR, still in the two-receiver case, numerical results show that RDIF achieves performance superior to optimal linear precoding and very close to the sum capacity.

We are currently working on extending the results of this paper for $K > 2$, motivated by the promising performance evidenced by numerical search.
Other potential extensions of this work include studying other objective functions such as weighted sum rate, other practical constraints such as a per-antenna power constraint, as well as the impact of imperfect CSIT.

\bibliographystyle{IEEEtran}
\bibliography{IEEEabrv,refs}

\begin{thebibliography}{10}
\providecommand{\url}[1]{#1}
\csname url@samestyle\endcsname
\providecommand{\newblock}{\relax}
\providecommand{\bibinfo}[2]{#2}
\providecommand{\BIBentrySTDinterwordspacing}{\spaceskip=0pt\relax}
\providecommand{\BIBentryALTinterwordstretchfactor}{4}
\providecommand{\BIBentryALTinterwordspacing}{\spaceskip=\fontdimen2\font plus
\BIBentryALTinterwordstretchfactor\fontdimen3\font minus
  \fontdimen4\font\relax}
\providecommand{\BIBforeignlanguage}[2]{{%
\expandafter\ifx\csname l@#1\endcsname\relax
\typeout{** WARNING: IEEEtran.bst: No hyphenation pattern has been}%
\typeout{** loaded for the language `#1'. Using the pattern for}%
\typeout{** the default language instead.}%
\else
\language=\csname l@#1\endcsname
\fi
#2}}
\providecommand{\BIBdecl}{\relax}
\BIBdecl

\bibitem{Tse.Viswanath}
D.~Tse and P.~Viswanath, \emph{Fundamentals of Wireless Communication}.\hskip
  1em plus 0.5em minus 0.4em\relax Cambridge University Press, 2005.

\bibitem{Caire2003}
G.~Caire and S.~Shamai, ``{On the achievable throughput of a multiantenna
  {G}aussian broadcast channel},'' \emph{{IEEE} Trans. Inf. Theory}, vol.~49,
  no.~7, pp. 1691--1706, 2003.

\bibitem{Viswanath2003}
P.~Viswanath and D.~Tse, ``Sum capacity of the vector {G}aussian broadcast
  channel and uplink-downlink duality,'' \emph{{IEEE} Trans. Inf. Theory},
  vol.~49, no.~8, pp. 1912--1921, Aug 2003.

\bibitem{Vishwanath.etal.2003.Duality-Achievable-Rates}
S.~{Vishwanath}, N.~{Jindal}, and A.~{Goldsmith}, ``Duality, achievable rates,
  and sum-rate capacity of {{Gaussian MIMO}} broadcast channels,'' \emph{{IEEE}
  Trans. Inf. Theory}, vol.~49, no.~10, pp. 2658--2668, Oct. 2003.

\bibitem{Yu.Cioffi.2004.Sum-Capacity-Gaussian}
W.~{Yu} and J.~{Cioffi}, ``Sum capacity of {{Gaussian}} vector broadcast
  channels,'' \emph{{IEEE} Trans. Inf. Theory}, vol.~50, no.~9, pp. 1875--1892,
  Sep. 2004.

\bibitem{Weingarten2006}
H.~Weingarten, Y.~Steinberg, and S.~Shamai, ``The capacity region of the
  {G}aussian multiple-input multiple-output broadcast channel,'' \emph{{IEEE}
  Trans. Inf. Theory}, vol.~52, no.~9, pp. 3936--3964, Sept 2006.

\bibitem{Yoo.Goldsmith.2006.Optimality-Multiantenna-Broadcast}
T.~{Yoo} and A.~{Goldsmith}, ``On the optimality of multiantenna broadcast
  scheduling using zero-forcing beamforming,'' \emph{{IEEE} J. Sel. Areas
  Commun.}, vol.~24, no.~3, pp. 528--541, Mar. 2006.

\bibitem{Bjornson2014}
E.~Bjornson, M.~Bengtsson, and B.~Ottersten, ``Optimal multiuser transmit
  beamforming: A difficult problem with a simple solution structure,''
  \emph{{IEEE} Signal Process. Mag.}, vol.~31, no.~4, pp. 142--148, jul 2014.

\bibitem{Lee2007}
J.~Lee and N.~Jindal, ``{High SNR Analysis for MIMO Broadcast Channels: Dirty
  Paper Coding Versus Linear Precoding},'' \emph{{IEEE} Trans. Inf. Theory},
  vol.~53, no.~12, pp. 4787--4792, dec 2007.

\bibitem{Hong2012}
S.-N. Hong and G.~Caire, ``Reverse compute and forward: A low-complexity
  architecture for downlink distributed antenna systems,'' in \emph{Proc. IEEE
  Int. Symp. Information Theory}, Jul. 2012, pp. 1147--1151.

\bibitem{Hong2013}
S.~N. Hong and G.~Caire, ``{Compute-and-forward strategies for cooperative
  distributed antenna systems},'' \emph{{IEEE} Trans. Inf. Theory}, vol.~59,
  no.~9, pp. 5227--5243, 2013.

\bibitem{He2014}
W.~He, B.~Nazer, and S.~Shamai, ``Uplink-downlink duality for
  integer-forcing,'' in \emph{Proc. IEEE Int. Symp. Information Theory}, Jun.
  2014, pp. 2544--2548.

\bibitem{He.etal.2014.Uplink-downlink-Duality-Integer-forcing:Iterative-Optimization}
------, ``Uplink-downlink duality for integer-forcing: Effective {SINRs} and
  iterative optimization,'' in \emph{Proc. IEEE Int. Workshop Signal Process.
  Adv. Wireless Commun. (SPAWC)}, Jun. 2014, pp. 474--478.

\bibitem{Nazer2011}
B.~Nazer and M.~Gastpar, ``Compute-and-forward: Harnessing interference through
  structured codes,'' \emph{{IEEE} Trans. Inf. Theory}, vol.~57, no.~10, pp.
  6463--6486, Oct. 2011.

\bibitem{Zhan2014}
J.~Zhan, B.~Nazer, U.~Erez, and M.~Gastpar, ``Integer-forcing linear
  receivers,'' \emph{{IEEE} Trans. Inf. Theory}, vol.~60, no.~12, pp.
  7661--7685, Dec 2014.

\bibitem{Ordentlich.Erez.2015.Precoded-Integer-Forcing-Universally}
O.~Ordentlich and U.~Erez, ``Precoded integer-forcing universally achieves the
  {MIMO} capacity to within a constant gap,'' \emph{{IEEE} Trans. Inf. Theory},
  vol.~61, no.~1, pp. 323--340, Jan. 2015.

\bibitem{Sakzad.Viterbo.2015.Full-Diversity-Unitary}
A.~Sakzad and E.~Viterbo, ``Full diversity unitary precoded integer-forcing,''
  \emph{{IEEE} Trans. Wireless Commun.}, vol.~14, no.~8, pp. 4316--4327, Aug.
  2015.

\bibitem{Yao.Wornell.2002.Lattice-reduction-aided-Detectors-MIMO}
H.~Yao and G.~W. Wornell, ``Lattice-reduction-aided detectors for {{MIMO}}
  communication systems,'' in \emph{Proc. IEEE Global Telecommunications
  Conference (GLOBECOM)}, vol.~1, Nov. 2002, pp. 424--428.

\bibitem{Fischer.etal.2016.Factorization-Approaches-Lattice-Reduction-Aided}
R.~Fischer, M.~Cyran, and S.~Stern, ``Factorization approaches in
  lattice-reduction-aided and integer-forcing equalization,'' in \emph{Int.
  Zurich Seminar on Communications}, Mar. 2016, pp. 108--112.

\bibitem{Windpassinger.etal.2004.Lattice-reduction-aided-Broadcast-Precoding}
C.~Windpassinger, R.~F.~H. Fischer, and J.~B. Huber, ``Lattice-reduction-aided
  broadcast precoding,'' \emph{{IEEE} Trans. Commun.}, vol.~52, no.~12, pp.
  2057--2060, Dec. 2004.

\bibitem{Stern.Fischer.2016.Advanced-Factorization-Strategies}
S.~Stern and R.~F.~H. Fischer, ``Advanced factorization strategies for
  lattice-reduction-aided preequalization,'' in \emph{Proc. IEEE Int. Symp.
  Information Theory}, Jul. 2016, pp. 1471--1475.

\bibitem{Zamir.2014.Lattice-Coding-Signals}
R.~Zamir, \emph{Lattice Coding for Signals and Networks}.\hskip 1em plus 0.5em
  minus 0.4em\relax Cambridge: {Cambridge University Press}, 2014.

\bibitem{Dummit.Foote}
D.~S. Dummit and R.~M. Foote, \emph{Abstract Algebra}, 3rd~ed.\hskip 1em plus
  0.5em minus 0.4em\relax John Wiley \& Sons, 2004.

\bibitem{Feng2013}
C.~Feng, D.~Silva, and F.~R. Kschischang, ``{An algebraic approach to
  physical-layer network coding},'' \emph{{IEEE} Trans. Inf. Theory}, vol.~59,
  no.~11, pp. 7576--7596, 2013.

\bibitem{Nazer.etal.2016.Expanding-Compute-and-Forward-Framework:}
B.~Nazer, V.~R. Cadambe, V.~Ntranos, and G.~Caire, ``Expanding the
  compute-and-forward framework: Unequal powers, signal levels, and multiple
  linear combinations,'' \emph{{IEEE} Trans. Inf. Theory}, vol.~62, no.~9, pp.
  4879--4909, Sep. 2016.

\bibitem{Erez.Zamir.2004.Achieving-1/2-Log}
U.~Erez and R.~Zamir, ``Achieving 1/2 log (1+{{SNR}}) on the {{AWGN}} channel
  with lattice encoding and decoding,'' \emph{{IEEE} Trans. Inf. Theory},
  vol.~50, no.~10, pp. 2293--2314, Oct. 2004.

\bibitem{Wiesel2008}
A.~Wiesel, Y.~Eldar, and S.~Shamai, ``Zero-forcing precoding and generalized
  inverses,'' \emph{{IEEE} Trans. Signal Process.}, vol.~56, no.~9, pp.
  4409--4418, sep 2008.

\bibitem{Micciancio.Goldwasser.2002.Complexity-Lattice-Problems}
D.~Micciancio and S.~Goldwasser, \emph{Complexity of {{Lattice
  Problems}}}.\hskip 1em plus 0.5em minus 0.4em\relax Boston, MA: {Springer
  US}, 2002.

\bibitem{Lenstra.etal.1982.Factoring-Polynomials-Rational}
A.~K. Lenstra, H.~W. Lenstra, and L.~Lov{\'a}sz,
  ``\BIBforeignlanguage{en}{Factoring polynomials with rational
  coefficients},'' \emph{\BIBforeignlanguage{en}{Mathematische Annalen}}, vol.
  261, no.~4, pp. 515--534, Dec. 1982.

\bibitem{Gan.etal.2009.Complex-Lattice-Reduction}
Y.~H. Gan, C.~Ling, and W.~H. Mow, ``Complex lattice reduction algorithm for
  low-complexity full-diversity mimo detection,'' \emph{{IEEE} Trans. Signal
  Process.}, vol.~57, no.~7, pp. 2701--2710, Jul. 2009.

\end{thebibliography}

\begin{IEEEbiography}[{\includegraphics[width=1in,height=1.25in,clip,keepaspectratio]{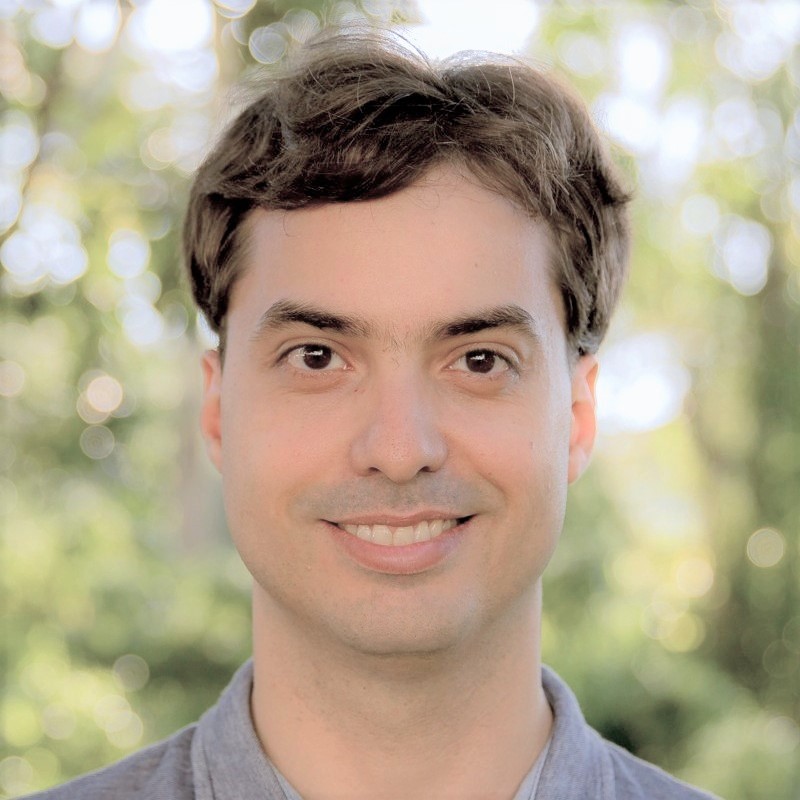}}]{Danilo Silva} (S'06--M'09) received the B.Sc.\ degree from the Federal University of Pernambuco (UFPE), Recife, Brazil, in 2002, the M.Sc.\ degree from the Pontifical Catholic University of Rio de Janeiro (PUC-Rio), Rio de Janeiro, Brazil, in 2005, and the Ph.D. degree from the University of Toronto, Toronto, Canada, in 2009, all in electrical engineering.
From 2009 to 2010, he was a Postdoctoral Fellow at the University of Toronto, at the \'Ecole Polytechnique F\'ed\'erale de Lausanne (EPFL), and at the State University of Campinas (UNICAMP).
In 2010, he joined the Department of Electrical Engineering, Federal University of Santa Catarina (UFSC), Brazil, where he is currently an Assistant Professor. His research interests include wireless communications, channel coding, information theory, and network coding.

Dr. Silva was a recipient of a CAPES Ph.D. Scholarship in 2005, the Shahid U. H. Qureshi Memorial Scholarship in 2009, and a FAPESP Postdoctoral Scholarship in 2010.
\end{IEEEbiography}

\begin{IEEEbiography}[{\includegraphics[width=1in,height=1.25in,clip,keepaspectratio]{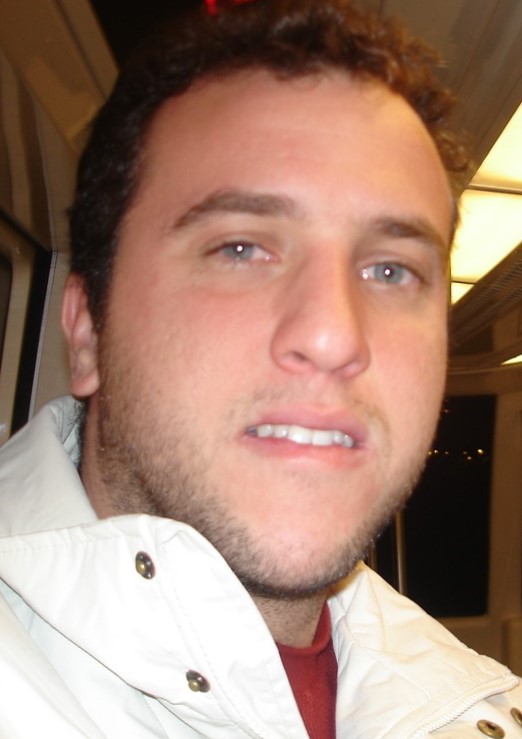}}]{Gabriel Pivaro} received his Teacher Certification in Mathematics from the São Paulo State University (UNESP) in 2005, M.Sc. and Ph.D. degrees in Electrical Engineering from University of São Paulo (USP) and State University of Campinas (UNICAMP), São Paulo, Brazil, in 2008 and 2016, respectively. 
From Apr. 2015 to Mar. 2016, he was a Research Scholar at Rice University, Houston, Texas. He is currently with the Radiocommunications Reference Center at the National Institute of Telecommunications (Inatel) Minas Gerais, Brazil, working with 5G networks, Channel Sounding and Cognitive Radio. From 1999 to 2012 he was working with Radio Access Network for Motorola, Nokia, Huawei, and Claro (América Móvil). Dr. Pivaro's research interests are 5G networks, Random Matrix Theory applied to MIMO communications, and Wireless Communications.
\end{IEEEbiography}

\begin{IEEEbiography}[{\includegraphics[width=1in,height=1.25in,clip,keepaspectratio]{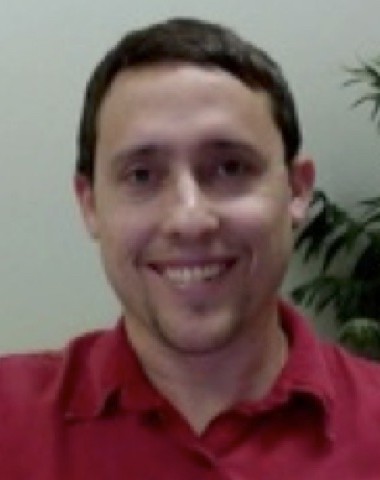}}]{Gustavo Fraidenraich} graduated in Electrical Engineering from the Federal University of Pernambuco, UFPE, Brazil, in 1997. He received his M.Sc. and Ph.D. degrees from the State University of Campinas, UNICAMP, Brazil, in 2002 and 2006, respectively. From 2006 to 2008, he worked as Postdoctoral Fellow at Stanford University (Star Lab Group) - USA. Currently, Dr. Fraidenraich is Assistant Professor at UNICAMP - Brazil and his research interests include Multiple Antenna Systems, Cooperative systems, Radar Systems and Wireless Communications in general.
\end{IEEEbiography}

\begin{IEEEbiography}[{\includegraphics[width=1in,height=1.25in,clip,keepaspectratio]{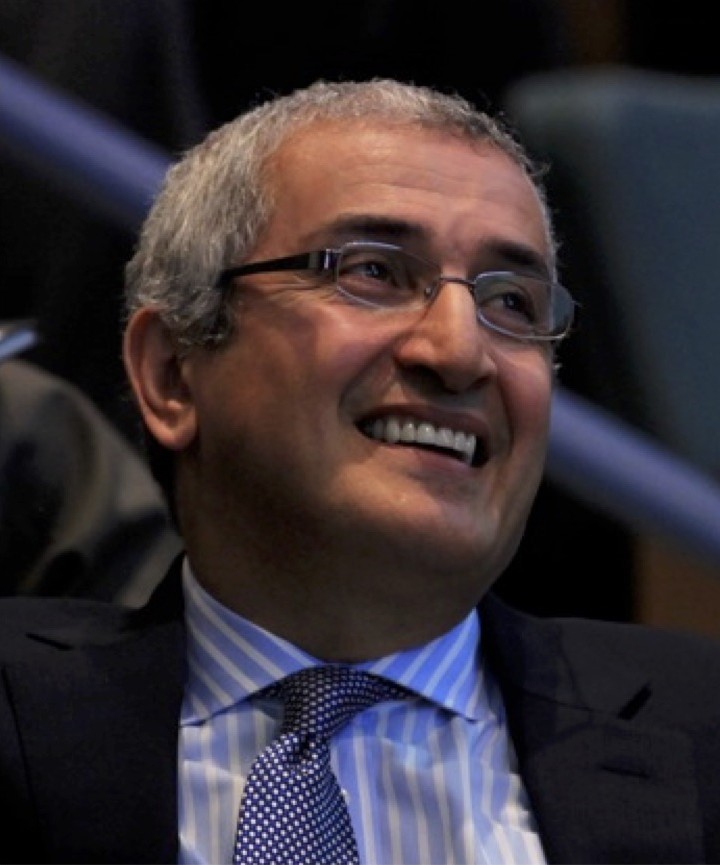}}]{Behnaam Aazhang} (S'82--M'82--SM'91--F'99) received his B.S. (with highest honors), M.S., and Ph.D. degrees in Electrical and Computer Engineering from University of Illinois at Urbana-Champaign in 1981, 1983, and 1986, respectively. From 1981 to 1985, he was a Research Assistant in the Coordinated Science Laboratory, University of Illinois. In August 1985, he joined the faculty of Rice University, Houston, Texas, where he is now the J.S. Abercrombie Professor in the Department of Electrical and Computer Engineering and a Professor and the Director of the Center on Neuro-Engineering, a multi-university research center in Houston,
Texas. From 2006 till 2014, he held an Academy of Finland Distinguished Visiting Professorship appointment (FiDiPro) at the University of Oulu, Oulu, Finland. Dr. Aazhang is a Fellow of IEEE and AAAS, a distinguished lecturer of IEEE Communication Society. He received an Honorary Doctorate degree from the University of Oulu, Finland (the highest honor that the university can bestow) in 2017 and IEEE ComSoc CTTC Outstanding Service Award ``For innovative leadership that elevated the success of the Communication Theory Workshop'' in 2016. He is a recipient of 2004 IEEE Communication Society's Stephen O. Rice best paper award for a paper with A. Sendonaris and E. Erkip. In addition, Sendonaris, Erkip, and Aazhang received IEEE Communication Society's 2013 Advances in Communication Award for the same paper. He has been listed in the Thomson-ISI Highly Cited Researchers and has been keynote and plenary speaker of several conferences. His research interests are signal processing, information theory, and their applications to neuroengineering and wireless communication and networks. Particular focus is on modeling neuronal circuits connectivity and the impact of learning on connectivity, on real-time closed-loop stabilization of neuronal systems to mitigate disorders such as epilepsy, Parkinson, depression, and obesity, on developing an understanding of cortical representation and fine-grained connectivity in the human language system, and on building microelectronics with large data analysis techniques to develop a fine-grained recording and modulation system to remediate language disorders.
\end{IEEEbiography}

\end{document}